%% file: main.tex
\documentclass[pdflatex,sn-basic]{sn-jnl}

\usepackage{graphicx}%
\usepackage{multirow}%
\usepackage{amsmath,amssymb,amsfonts}%
\usepackage{amsthm}%
\usepackage{mathrsfs}%
\usepackage[title]{appendix}%
\usepackage{textcomp}%
\usepackage{manyfoot}%
\usepackage{booktabs}%
\usepackage{algorithm}%
\usepackage{algorithmicx}%
\usepackage{algpseudocode}%
\usepackage{listings}%

\usepackage{acronym}
\usepackage{comment}
\usepackage{subcaption}
\usepackage[x11names]{xcolor}
\usepackage{xspace}

\usepackage{tikz}
\usetikzlibrary{calc}
\usetikzlibrary{fit}
\usetikzlibrary{positioning}
\usetikzlibrary{shapes.multipart}
\usetikzlibrary{shapes.geometric}
\usetikzlibrary{matrix}
\usetikzlibrary{backgrounds}
\usetikzlibrary{patterns}
\usetikzlibrary{patterns.meta}
\usetikzlibrary{arrows.meta}

\newcommand{\idest}{i.e.,\xspace}
\newcommand{\eg}{e.g.,\xspace}

\newcommand{\wrt}{w.r.t.\xspace}
\newcommand{\rr}{round-robin\xspace}

\newcommand{\secref}[1]{Section \ref{#1}\xspace}
\newcommand{\figref}[1]{Figure \ref{#1}\xspace}
\newcommand{\tabref}[1]{Table \ref{#1}\xspace}

\theoremstyle{thmstyleone}%
\theoremstyle{thmstyletwo}%

\theoremstyle{thmstylethree}%

\input{acronyms}

\begin{document}

\title[Minor Embedding for Quantum Annealing with Reinforcement Learning]{Minor Embedding for Quantum Annealing with Reinforcement Learning}


\author*[1]{\fnm{Riccardo} \sur{Nembrini}}\email{riccardo.nembrini@polimi.it}

\author*[1]{\fnm{Maurizio} \sur{Ferrari Dacrema}}\email{maurizio.ferrari@polimi.it}

\author[1]{\fnm{Paolo} \sur{Cremonesi}}\email{paolo.cremonesi@polimi.it}

\affil[1]{\orgname{Politecnico di Milano}, \city{Milano}, \country{Italy}}


\abstract{Quantum Annealing (QA) is a quantum computing paradigm for solving combinatorial optimization problems formulated as Quadratic Unconstrained Binary Optimization (QUBO) problems. An essential step in QA is minor embedding, which maps the problem graph onto the sparse topology of the quantum processor. This process is computationally expensive and scales poorly with increasing problem size and hardware complexity. Existing heuristics are often developed for specific problem graphs or hardware topologies and are difficult to generalize. Reinforcement Learning (RL) offers a promising alternative by treating minor embedding as a sequential decision-making problem, where an agent learns to construct minor embeddings by iteratively mapping the problem variables to the hardware qubits. We propose a RL-based approach to minor embedding using a Proximal Policy Optimization agent, testing its ability to embed both fully connected and randomly generated problem graphs on two hardware topologies, Chimera and Zephyr. The results show that our agent consistently produces valid minor embeddings, with reasonably efficient number of qubits, in particular on the more modern Zephyr topology. Our proposed approach is also able to scale to moderate problem sizes and adapts well to different graph structures, highlighting RL's potential as a flexible and general-purpose framework for minor embedding in QA.
}

\keywords{Quantum Annealing, Minor Embedding, Reinforcement Learning, Proximal Policy Optimization}



\maketitle

\section{Introduction}
\label{introduction}
Among the different paradigms of Quantum Computation, Quantum Annealing (QA) operates by representing an optimization problem as an energy minimization one, and then evolving a physical quantum system from an initial default configuration towards a final one that is constructed based on the target problem \citep{PhysRevE.58.5355,johnson2011quantum}. QA is often used to solve combinatorial optimization problems formulated as Quadratic Unconstrained Binary Optimization (QUBO) ones. While the QUBO formulation is very general, the physical hardware of a \ac{QPU}, or Quantum Annealer, exhibits a specific topology and connections between the physical qubits. Typical optimization problems in machine learning and other domains are not natively compatible with the hardware connectivity and must be transformed into other equivalent ones having a structure that is compatible with the QPU, in a process called \acfi{ME}.

Minor embedding is, however, an optimization problem of its own and often acts as a computational bottleneck \citep{DBLP:conf/sigir/DacremaMN0FC22}, requiring a time that exceeds by orders of magnitude the actual quantum annealing process. Furthermore, existing heuristics lack flexibility because they do not include objective functions that can be adjusted according to the scenario of interest. While their computational cost could be mitigated by pre-computing minor embeddings or developing ad-hoc heuristics for fixed problem graphs, such as fully connected graphs, on specific hardware topologies \citep{DBLP:journals/qip/BoothbyKR16}, this strategy is limited by the large number of possible problem graphs and hardware variations due to inactive qubits or differing topological layouts. Moreover, the minor's structure can significantly affect solution quality. For example, longer qubit chains increase the likelihood of errors and inconsistencies during annealing, as well as make more difficult for the qubits to change their state and therefore optimize the desired objective function. The relationship between minor embedding structure and solution quality has not yet been fully understood \citep{vinci2015quantum,DBLP:journals/qmi/PelliniD24}.

These limitations motivate the search for alternative more flexible machine learning approaches to minor embedding, that would also allow the freedom to define new objective functions. In this work, we explore the potential of Reinforcement Learning (RL), to build minor embeddings as a sequential decision-making problem. While RL methods bring a set of challenges of their own, \eg long training time and instabilities, they also provide a much higher degree of flexibility and adaptability to changing conditions, which can support their generalization across problem instances and hardware topologies.

In summary, the contributions of this work are as follows:
\begin{itemize}
    \item We approach minor embedding as a sequential decision-making problem and propose a Proximal Policy Optimization (PPO) based agent to generate minor embeddings.
    \item To improve learning efficiency and exploit the inherent symmetries of the hardware topology, we propose a set of data augmentation strategies that enhance generalization and policy robustness in particular on randomly generated problem graphs.
    \item We conduct a detailed comparison of minor embedding quality, success rate, and qubit efficiency across two widely used quantum hardware topologies, Chimera and Zephyr, highlighting the differences in agent performance under varying connectivity constraints.
\end{itemize}

The methods, experiments, and results presented in this paper are an extension of our prior work in \citep{nembrini_me}.
Our results show that RL agents can produce valid and efficient minor embeddings, particularly on modern topologies such as Zephyr. These findings highlight the potential of Reinforcement Learning as a flexible and extensible framework for addressing minor embedding in QA, as well as some limitations of the proposed method which suggest possible directions for future work.

\section{Background}
\label{sec:bg}
This section provides the necessary background on the three key topics: the computational paradigm of Quantum Annealing, the Minor Embedding problem, and the Reinforcement Learning paradigm which our proposed method is based on.

\subsection{Quantum Annealing}
\label{sec:bg:qa}
Quantum Annealing (QA) is a metaheuristic quantum algorithm designed to solve combinatorial optimization problems by exploiting the principles of quantum mechanics \citep{PhysRevE.58.5355,farhi2000quantum,johnson2011quantum}. The approach relies on formulating the optimization problem as an energy minimization of a real physical system, encoding it as a Hamiltonian \( H_P \) (known as the problem Hamiltonian) whose ground state corresponds to the optimal solution.

QA operates by initializing the quantum system in the ground state of a simple Hamiltonian \( H_0 \), for which the ground state is easy to prepare, typically an equal superposition. Over the course of the annealing schedule, the system evolves according to a time-dependent Hamiltonian of the form:
\begin{equation}
    H(t) = (1 - s(t)) H_0 + s(t) H_P,
\end{equation}
where \( t \in [0, T] \) is the time parameter, \( T \) is the total annealing time, and \( s(t) \) is a monotonically increasing function that controls the evolution schedule satisfying \( s(0) = 0 \) and \( s(T) = 1 \). If the evolution is sufficiently slow, according to the requirements of the adiabatic theorem \citep{born1928beweis}, the system will remain in its instantaneous ground state. At the end of the evolution, when $H(T) = H_P$, the ground state of the system will also correspond to the ground state of $H_P$ hence to the solution of the problem.

The problem Hamiltonian \( H_P \) is typically constructed from a QUBO (Quadratic Unconstrained Binary Optimization) formulation:
\begin{equation}
    \label{eq:qubo}
    \min_{x \in \{0,1\}^n} x^\top Q x,
\end{equation}
where \( Q \in \mathbb{R}^{n \times n} \) is a symmetric matrix defining the cost function. This classical problem is mapped onto an Ising Hamiltonian of the form:
\begin{equation}
    \label{eq:ising}
    H_{\text{Ising}} = -\sum_{i=1}^{n} h_i s_i - \sum_{i=1}^{n}\sum_{j=i+1}^{n} J_{ij} s_i s_j,
\end{equation}
where $s_i \in \{-1, +1\}$ are spin variables representing the state of qubit \( i \), \( h_i \in \mathbb{R} \) are linear biases, and \( J_{ij} \in \mathbb{R} \) are quadratic couplings between qubits. By looking at Eq. \ref{eq:qubo} and \ref{eq:ising} one can see that there is a biunivocal correspondence between the two. Hence, by constructing a quantum system that implements and minimizes an Ising Hamiltonian, one can also minimize QUBO problems. This approach has gained popularity because it allows to formulate many difficult problems rather easily \citep{lucas2014ising,Glover2022}. Many applications of QA have been proposed in the fields of machine learning \citep{Neven2009, Mandra2016, OMalleyVAA17, Mott2017, KumarBTD18, ottaviani2018low, Neukart2018_cluster, Neukart2018, WillschWRM20, nembrini_cqfs, DBLP:conf/recsys/NembriniCDC22, DBLP:conf/sigir/DacremaMN0FC22, DBLP:conf/clef/PasinDCF24, DBLP:conf/qce/CarugnoDC24}, chemistry \citep{HernandezA17, Xia2018, streif2019solving, Micheletti2021}, as well as logistics and optimization \citep{RieffelVODPS15, StollenwerkLJ17, Ikeda2019, Ohzeki2020, Carugno2022, DBLP:conf/qce/ChiavassaMPDC22}.

\begin{figure}
     \centering
     \begin{subfigure}[b]{0.3\textwidth}
         \centering
         \includegraphics[width=\textwidth]{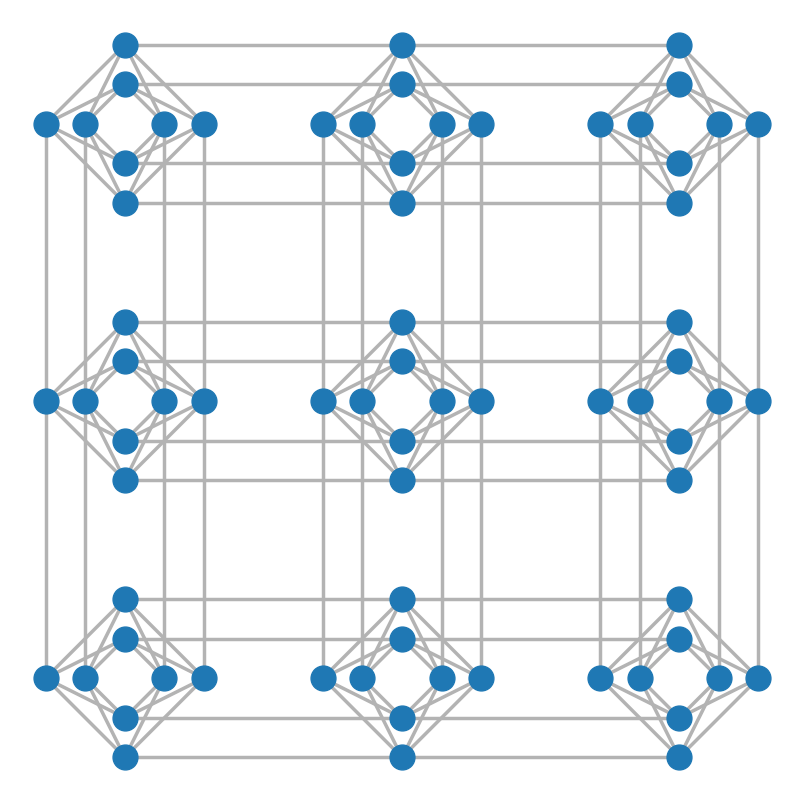}
         \caption{Chimera}
         \label{fig:topology:chimera}
     \end{subfigure}
     \hfill
     \begin{subfigure}[b]{0.3\textwidth}
         \centering
         \includegraphics[width=\textwidth]{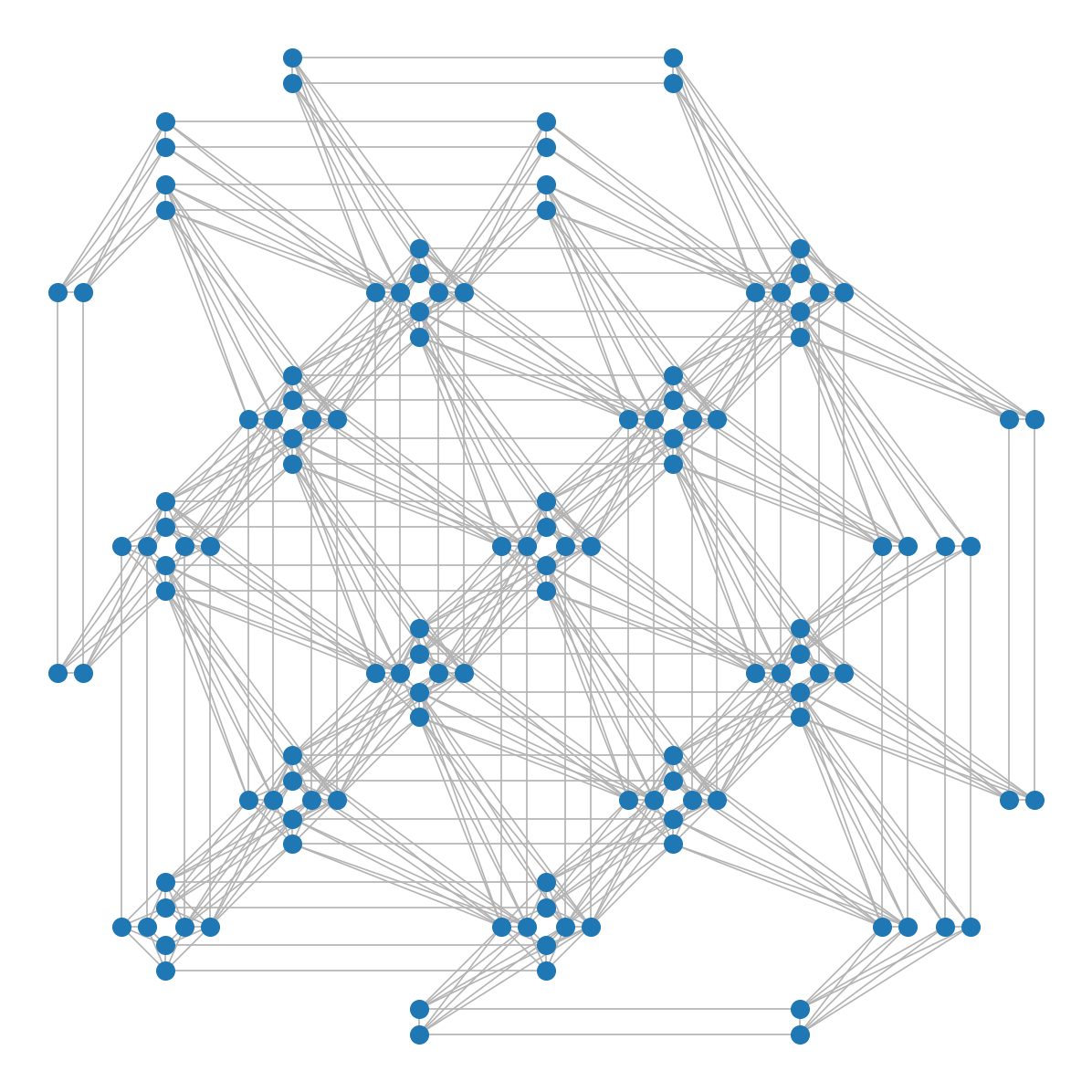}
         \caption{Pegasus}
         \label{fig:topology:pegasus}
     \end{subfigure}
     \hfill
     \begin{subfigure}[b]{0.3\textwidth}
         \centering
         \includegraphics[width=\textwidth]{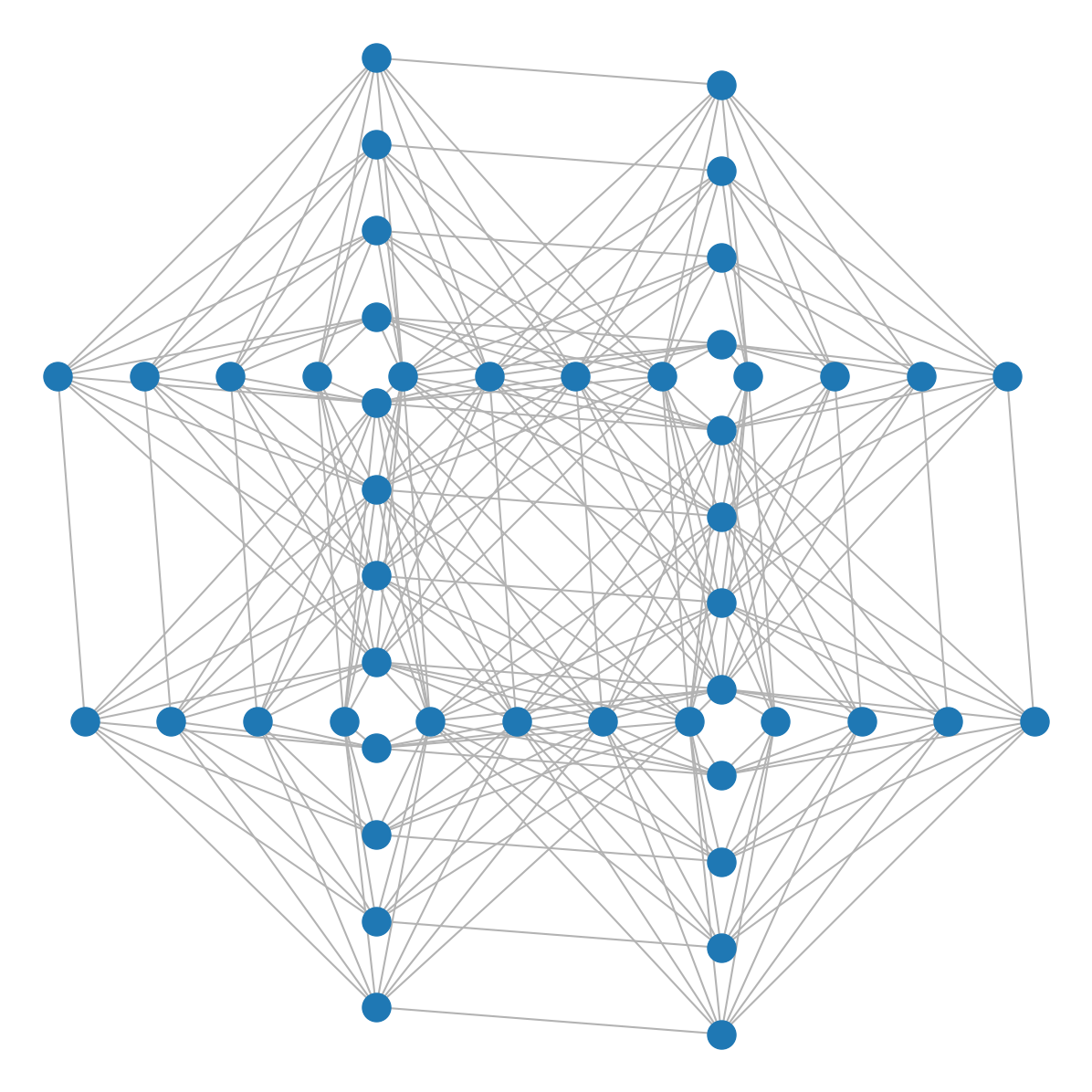}
         \caption{Zephyr}
         \label{fig:topology:zephyr}
     \end{subfigure}
    \caption{Portions of the currently existing topologies of physical quantum annealers produced by D-Wave, from oldest (Chimera) to most recent (Zephyr). The main difference between them is the number of connections between qubits, up to 6 in Chimera, up to 15 in Pegasus and up to 20 in Zephyr.}
    \label{fig:topology}
\end{figure}

Despite its general formulation, in practice QA is implemented on hardware platforms such as D-Wave quantum processors, where the qubits are laid out in sparsely connected topologies, see \figref{fig:topology}.
These topologies are typically composed of a repeated grid of small subgraphs called \emph{unit cells}, whose structure depends on the topology.
Given that the optimization problem must be physically encoded into the hardware, there must be a biunivocal correspondence between the two. This is obtained via the \emph{Minor Embedding}, in which the optimization problem is first represented as a graph, where problem variables that have a non-zero quadratic coefficient are connected, and then it is transformed in a new and equivalent one which can be directly implemented on the physical hardware.

\subsection{Minor Embedding}
\label{sec:bg:me}
When a problem is expressed in the QUBO formalism, it is typically assumed that all problem variables can interact freely. However, executing such a problem on a quantum annealer requires mapping it to the physical hardware, which imposes strict topological constraints. Each qubit in the quantum processing unit (\ac{QPU}) can only interact with a limited set of neighbouring qubits, determined by the architecture (e.g., Chimera or Zephyr), see Figure \ref{fig:topology}. As a result, the problem must be transformed to conform to the physical connectivity of the device, a process known as \emph{Minor Embedding} (ME).

The central idea of ME is to first represent the optimization problem as a graph, where the variables that are associated to a quadratic coefficient are connected. Then, if a problem variable must be connected to more other variables than the physical connectivity of the hardware allows, that variable will be represented using multiple physical qubits, \idest a \emph{chain}, and the connections will be split among them. 
The chains form connected sub-graphs within the hardware topology and are forced to act coherently by applying strong coupling coefficients ($J_{ij}$) between their qubits. These couplings encourage all qubits in a chain to be in the same state during the annealing process.

Formally, ME involves two main steps: \emph{node mapping}, where each node of the problem graph \( G \) is assigned to one or more physical qubits in the hardware graph \( H \), and \emph{parameter setting}, which distributes the bias ($h_i$) and coupling ($J_ij$) values among the embedded variables. The resulting embedded graph \( G_{\text{emb}} \subseteq H \), also called \emph{minor}, contains intra-chain edges (within chains) and inter-chain edges (between different chains). The minor embedding is \emph{valid} if contracting all intra-chain edges in \( G_{\text{emb}} \) recovers the original problem graph \( G \).

The embedded Ising energy function can be written as:
\begin{equation}
    \label{eq:me}
    \mathcal{E} = \sum_{i \in V(G)}\left(\sum_{i_k \in V(C_i)}h^\prime_{i_k}s_{i_k} + \sum_{i_p,i_q \in E(C_i)} F^{pq}_is_{i_p}s_{i_q} \right) + \sum_{i, j \in E(G) \setminus E(C)}J_{ij}s_is_j,
\end{equation}
where $V(G)$ is the set of nodes in graph $G$, $E(G)$ is the set of edges in graph $G$, \( C_i \) denotes the chain assigned to problem variable \( i \), and \( F^{pq}_i < 0 \) are the intra-chain couplings. The original bias \( h_i \) is partitioned among the qubits in the chain such that \( \sum_{i_k \in V(C_i)} h^\prime_{i_k} = h_i \), while inter-chain couplings \( J_{ij} \) remain unchanged.

Minor embedding quality plays a crucial role in the effectiveness of quantum annealing. Long chains are more prone to \emph{chain breaks}, which occur when qubits within a chain do not reach the same final state, meaning that the quantum annealer has moved away from the ground state. Chain breaks can yield suboptimal or unfeasible solutions, where constraints are not met. Therefore, minimizing chain length, maintaining compact minors, and tuning chain strengths appropriately are essential objectives in the minor embedding process.

In light of this it is clear how minor embedding is NP-hard, and that exact solutions are computationally intractable in the general case, for this reason various heuristic methods have been proposed \citep{DBLP:journals/corr/CaiMR14,DBLP:journals/qip/BoothbyKR16,fang2020minimizing,bernal2020integer,pelofske20244}. One of the most commonly used tools is \texttt{minorminer} \citep{DBLP:journals/corr/CaiMR14}\footnote{\url{https://github.com/dwavesystems/minorminer}}, developed by D-Wave, which as we will describe implements a stochastic algorithm to construct a valid minor of a problem graph into a hardware graph like Chimera or Zephyr. However, these methods are often developed for specific hardware or graph topologies and, being heuristics, do not allow to control the actual optimization objective.

\subsection{\texttt{minorminer}}
One of the most commonly used tools for ME is \texttt{minorminer} \citep{DBLP:journals/corr/CaiMR14}, developed by D-Wave, which implements a stochastic algorithm to construct a minor embedding of a problem graph into a hardware graph like Chimera or Zephyr.

The basic idea is to embed one problem variable at a time. \texttt{minorminer} selects a not-yet-embedded problem variable and tries to assign it to a chain of physical qubits that would allow it to reach the other problem variables it must be connected with, according to the problem graph. If a suitable chain is found, the variable is added to the minor; otherwise, the algorithm can backtrack, removing previously assigned chains and trying different paths.
To do this, \texttt{minorminer} uses a greedy and randomized strategy:
\begin{itemize}
    \item Problem variables are considered in an order that prioritizes high connectivity or other heuristics.
    \item For each variable, it attempts to grow a chain using shortest paths to already embedded neighbors.
    \item If no valid chain is found, it probabilistically selects a different placement or backtracks.
\end{itemize}

\begin{figure}
    \centering
    \begin{subfigure}[b]{0.25\textwidth}
        \centering
        \includegraphics[width=\textwidth]{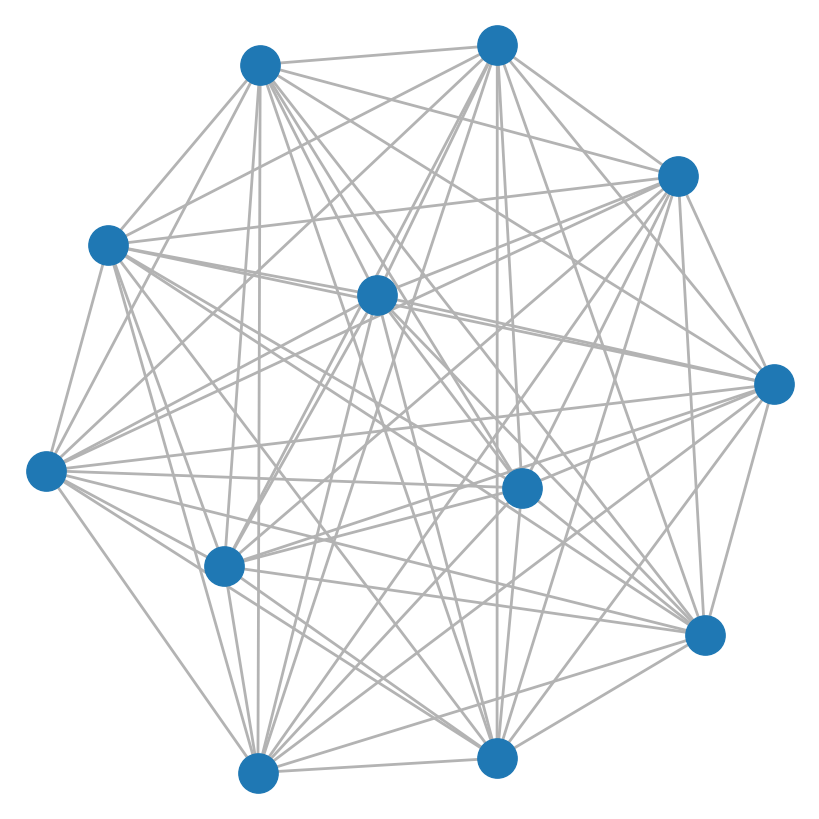}
        \caption{Problem graph G}
        \label{fig:minor:graph}
    \end{subfigure}
    \\
    \hfill
    \begin{subfigure}[b]{0.25\textwidth}
        \centering
        \includegraphics[width=\textwidth]{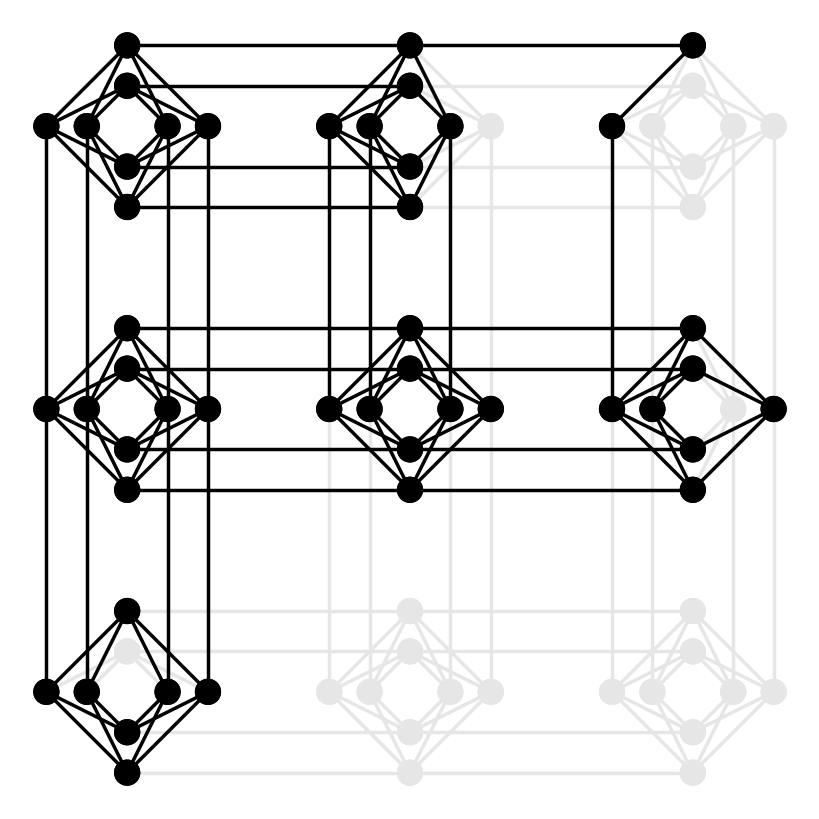}
        \caption{Chimera (47 nodes)}
        \label{fig:minor:chimera}
    \end{subfigure}
    \hfill
    \begin{subfigure}[b]{0.25\textwidth}
        \centering
        \includegraphics[width=\textwidth]{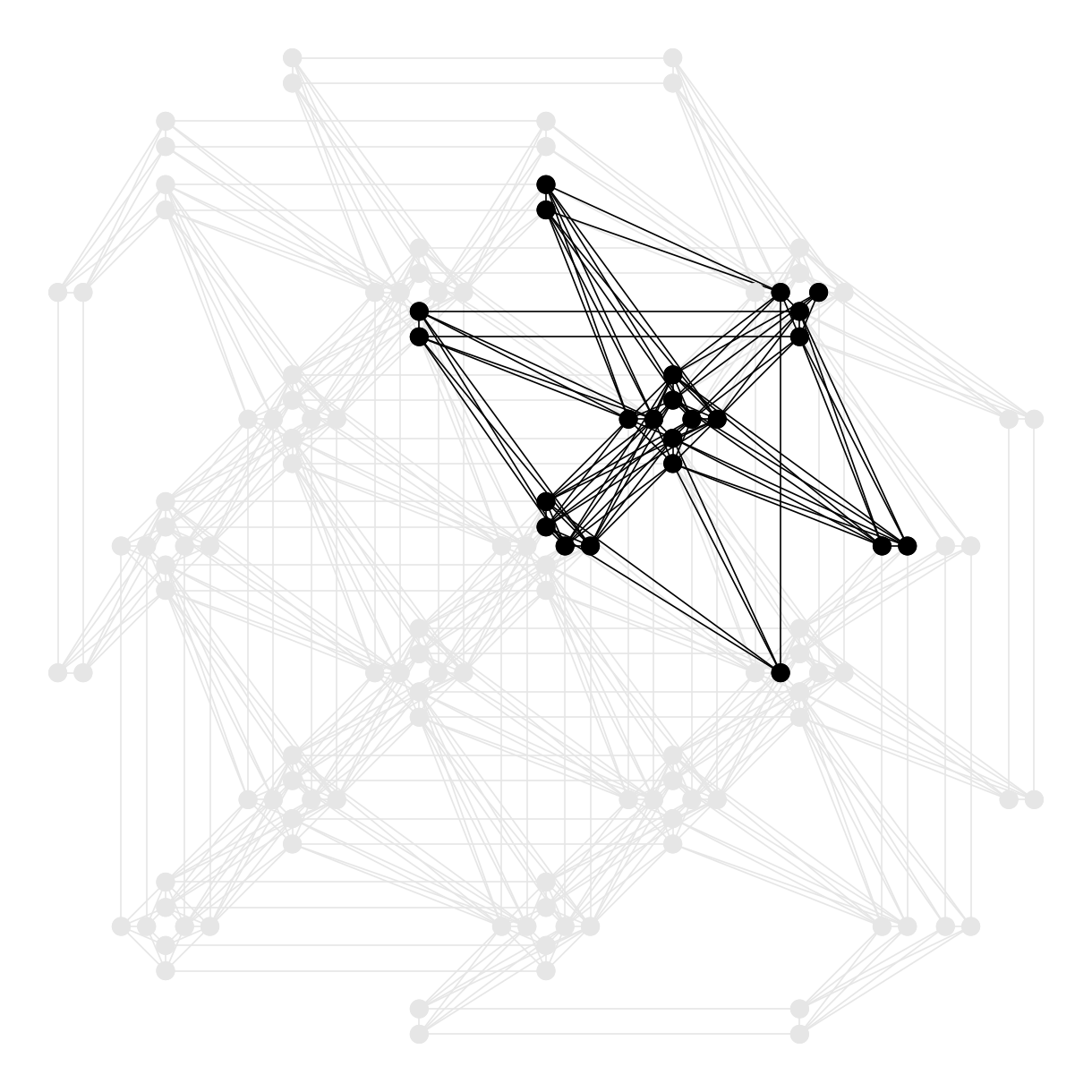}
        \caption{Pegasus (23 nodes)}
        \label{fig:minor:pegasus}
    \end{subfigure}
    \hfill
    \begin{subfigure}[b]{0.25\textwidth}
        \centering
        \includegraphics[width=\textwidth]{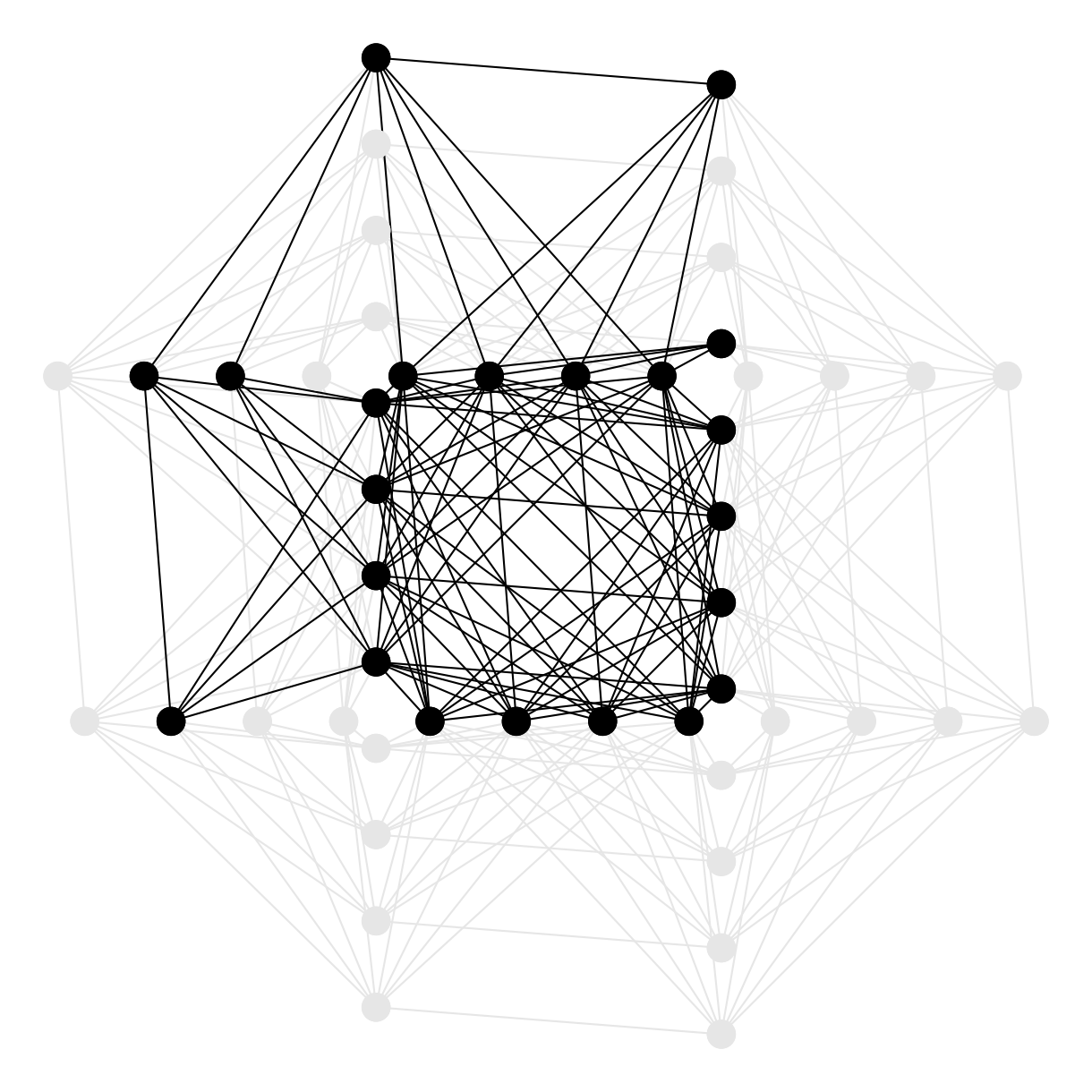}
        \caption{Zephyr (22 nodes)}
        \label{fig:minor:zephyr}
    \end{subfigure}
    \caption{Example of node mappings from a fully connected graph G of 12 nodes on different QA hardware topologies, obtained with \texttt{minorminer}. Highlighted in \ref{fig:minor:chimera}, \ref{fig:minor:pegasus} and \ref{fig:minor:zephyr} are only the nodes and edges that are part of the minor embeddings (respectively 47, 23 and 22 nodes). The chains are not differentiated for readability.}
    \label{fig:minor}
\end{figure}

This makes \texttt{minorminer} able to escape from local failures. The process continues until all problem variables are embedded, or a timeout or failure condition is reached. The output is a mapping from problem variables to chains of physical qubits. An example of minor embedding of a fully connected graph on different hardware topologies is reported in \figref{fig:minor}. 
\texttt{minorminer} does not guarantee optimality, and it is sensitive to the ordering of the problem variables and to random choices during search. For this reason, the minor embedding varies from one run to another. In practice, it is often run multiple times to find better minors, especially for dense problem graphs.

The main strengths of \texttt{minorminer} are its ability to handle large and irregular graphs, and its efficiency on modern hardware topologies it is optimized for. However, the time required to find a minor embedding will still be substantial compared to the quantum annealing time, the minor embeddings it finds will still use long chains, depending on the problem, and it does not allow easy customization to tailor its behaviour to specific goals. This motivates the exploration of alternative approaches based on machine learning.

\subsection{Reinforcement Learning}
\label{sec:bg:rl}

\begin{figure}
    \centering
    \begin{tikzpicture}[
        box/.style={rectangle, draw=black, thick, minimum width=4mm, minimum height=4mm},
        ]
        \node[box](agent){Agent};
        \node[box](env)[right=of agent]{Environment};
        
        \draw[dashed,->] ([yshift=-0.92cm]$([xshift=5]env.south)!0.8!([xshift=-7]agent.south)$) .. controls +(left:0) and +(down:0.75) .. ([xshift=-7]agent.south)
            node[pos=0.7,left]{$s_t$};
        
        \draw[->] (agent.north) .. controls +(up:1.2) and +(up:1.2) .. (env.north)
            node[pos=0.5,below]{$a_t$};
        
        \draw[->] (env.south) .. controls +(down:1.2) and +(down:1.2) .. (agent.south)
            node[pos=0.5,above]{$s_{t+1}$}
            node[pos=0.5,below]{$r_t$};
    \end{tikzpicture}
    \caption{The fundamental interaction loop of RL. At time step $t$ the agent, after observing the environment's state $s_t$, performs actions $a_t$. The environment reacts to such action by giving reward $r_t$ to the agent and then transitioning to state $s_{t+1}$. The new state will be subsequently observed by the agent at the next time step.
    }
    \label{fig:rl:int}
\end{figure}
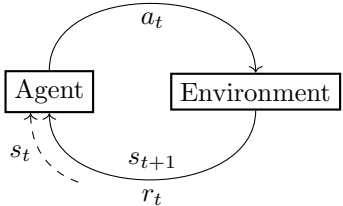

\acfi{RL} is a machine learning paradigm that involves an agent  iteratively interacting with an environment described by a Markov Decision Process (MDP) \citep{DBLP:books/lib/SuttonB2018}.
At each discrete time step $t$, the agent observes a state $s_t$ of the environment and selects an action $a_t$ based on this observation.
The chosen action influences the environment, causing a transition to a subsequent state $s_{t+1}$ according to the dynamics defined by the underlying MDP, and results in receiving a scalar reward signal $r_{t}$.
This interaction is cyclically repeated, as represented in \figref{fig:rl:int}, until a termination condition is reached at the final step $T$.
At that point, the agent has completed an \emph{episode}, collecting a number of rewards.
The goal of the agent is to maximize the expected cumulative reward, also known as the \emph{return}, obtained during each episode and defined as $G_t = \sum_{k=0}^{T-t} \gamma^k r_{t+k+1}$, where $\gamma \in [0,1]$ is a discount factor that balances the importance of immediate versus future rewards.
Maximizing this return allows the agent to learn an optimal strategy, called \emph{policy}, for navigating the state-action space.
Episodes are thus repeated to allow the agent to systematically explore diverse state-action configurations, thereby enabling learning of effective strategies.

As an illustrative example, consider a robotic agent tasked with traversing a field containing obstacles.
In such a scenario, the state observation might include the robot's current position within the field and the locations of nearby obstacles.
Available actions could consist of discrete movements, such as steps in one of the four cardinal directions.
Following each action, the agent receives a numerical reward reflecting its effectiveness; for instance, a small negative reward for each move could incentivize efficient paths by penalizing unnecessary steps and a large negative reward could discourage attempting to move in a direction that is obstructed.
Upon execution of an action, the environment transitions to a new state corresponding to the updated position of the robot.
An episode concludes upon reaching a specified destination or encountering a termination condition (e.g., collision with an obstacle or successful traversal).
By repeating episodes under varying conditions, the agent progressively learns improved strategies to navigate the environment effectively.

This framework exemplifies the fundamental reinforcement learning paradigm adopted throughout this work.
The training algorithm is introduced in the next sections.

\subsubsection{Actor-Critic Methods}
\label{sec:actor_critic}

Actor-critic methods constitute a prominent class of reinforcement learning algorithms characterized by employing two complementary structures: the \emph{actor}, which selects actions based on the current policy, and the \emph{critic}, which estimates the value associated with states or state-action pairs \citep{DBLP:books/lib/SuttonB2018, DBLP:conf/nips/KondaT99, DBLP:conf/icml/MnihBMGLHSK16}.
Within this framework, the policy $\pi$ is formally defined as a probability distribution over possible actions given a state:
\begin{equation}
    \label{eq:policy}
    \pi(a|s) = P(a_t = a\ |\ s_t = s).
\end{equation}

The agent aims to find an optimal policy $\pi^*$, which maximizes the expected cumulative reward, known as the \emph{return}, defined as:
\begin{equation}
    G_t = \sum_{k=0}^{T - t} \gamma^k r_{t+k+1},
\end{equation}
where $\gamma \in [0,1]$ is a discount factor balancing immediate versus future rewards, and $T$ denotes the terminal step of the episode.

In actor-critic methods, the actor directly represents the policy, \idest the learned model that chooses the actions, and is parameterized by a set of parameters $\theta$ as $\pi^{\theta}(a|s)$, while the critic estimates the expected returns associated with states or state-action pairs through a separate parameterized structure, typically denoted by parameters $\phi$.
Specifically, the critic approximates the \emph{value function}:
\begin{equation}
    \label{eq:value_function}
    v_\pi(s) = \mathbb{E}_\pi[G_t\ |\ S_t = s],
\end{equation}
which quantifies the expected cumulative reward starting from a given state $s$ and following the current policy $\pi$ thereafter.
Alternatively, the critic may approximate the \emph{state-action value function}:
\begin{equation}
    q_\pi(s, a) = \mathbb{E}_\pi[G_t\ |\ S_t = s, A_t = a],
\end{equation}
representing the expected cumulative reward obtained by taking action $a$ from state $s$ and subsequently following policy $\pi$.

By comparing the state-action value function and the value function, actor-critic methods estimate the \emph{advantage function}:
\begin{equation}
\label{eq:advantage}
    A_\pi(s,a) = q_\pi(s,a) - v_\pi(s),
\end{equation}
which measures how favourable selecting action $a$ in state $s$ is relative to the average action selection dictated by policy $\pi$ at that state.
Actions yielding positive advantage values are thereby reinforced, whereas actions with negative advantages are discouraged, guiding the policy toward optimal behaviour.
Actor-critic methods thus explicitly maintain separate policy and value estimators in order to improve stability, especially in environments characterized by continuous or high-dimensional action spaces \citep{DBLP:conf/nips/KondaT99,DBLP:journals/tsmc/GrondmanBLB12}.

The subsequent section details the Proximal Policy Optimization (PPO) algorithm, an actor-critic method specifically adopted for training the agents in this work.

\subsubsection{Proximal Policy Optimization}
\label{sec:rl:ppo}

\acfi{PPO} is an actor-critic algorithm proposed by~\cite{DBLP:journals/corr/SchulmanWDRK17}, specifically designed to overcome some of the stability and efficiency issues that affected earlier policy gradient methods.
PPO aims at improving policy optimization performance while maintaining computational simplicity, achieving a beneficial balance between robustness and ease of implementation.

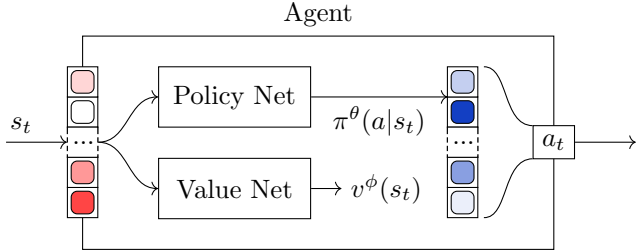
\begin{figure}
    \centering
    \begin{tikzpicture}[
            box/.style={minimum width=4mm, minimum height=4mm},
            cell/.style={minimum width=3mm, minimum height=3mm},
            network/.style={minimum width=20mm, minimum height=8mm},
            line/.style={solid},
            dline/.style={densely dashed},
        ]

        \coordinate (base) at (-20mm,-14mm);
        \coordinate (actions) at (30mm,-14mm);
        \coordinate (action) at (42mm,-6mm);
        \coordinate (fillpad) at ($(base) + (4mm,0)$);
        \coordinate (actionpad) at ($(action) + (-4mm,0)$);

        \node[network,draw] (pn) at (0,0) {Policy Net};
        \node[network,draw,below=4mm of pn] (vn) {Value Net};

        \node[fit=(pn)(vn)(fillpad)(actions)(actionpad),inner sep=4mm,draw] (agent) {};
        \node[above=0mm of agent]{Agent};

        \foreach [count=\i from 0,evaluate=\i as \y using \i*4] \shade in {90, 50, 30, 0, 20}{
            \node [box,draw,fill=white] (s\i) at ($(0,\y mm) + (base)$) {};
            \node [cell,draw,fill=Firebrick1!\shade!white,rounded corners=0.8mm] (c\i) at ($(0,\y mm) + (base)$) {};
        }
        \node[box,draw,fill=white,text width=2.7mm,inner sep=0pt] (etc) at ($(0,8mm) + (base)$) {...};
        \draw[white, very thick, dotted] ($(-2mm,6.4mm) + (base)$) -- ($(-2mm,9.6mm) + (base)$) ($(2mm,6.4mm) + (base)$) -- ($(2mm,9.6mm) + (base)$);
        
        \draw[->] ([xshift=-8mm]etc.west) -- (etc.west)
            node[near start,above]{$s_t$};

        \draw[->] (etc.east) .. controls ($(4mm,0) + (etc.east)$) and ($(-4mm,0) + (pn.west)$) .. (pn.west) {};
        \draw[->] (etc.east) .. controls ($(4mm,0) + (etc.east)$) and ($(-4mm,0) + (vn.west)$) .. (vn.west) {};

        \foreach [count=\i from 0,evaluate=\i as \y using \i*4] \shade in {10, 60, 30, 120, 30}{
            \node [box,draw,fill=white] (a\i) at ($(0,\y mm) + (actions)$) {};
            \node [cell,draw,fill=RoyalBlue3!\shade!white,rounded corners=0.8mm] (ca\i) at ($(0,\y mm) + (actions)$) {};
        }
        \node[box,draw,fill=white,text width=2.7mm,inner sep=0pt] at ($(0,8mm) + (actions)$) {...};
        \draw[white, very thick, dotted] ($(-2mm,6.4mm) + (actions)$) -- ($(-2mm,9.6mm) + (actions)$) ($(2mm,6.4mm) + (actions)$) -- ($(2mm,9.6mm) + (actions)$);

        \draw[->] (pn.east) -- ([xshift=-2mm,yshift=14mm]actions.west) {}
            node[below,midway] {$\pi^\theta(a|s_t)$};

        \node[right=4mm of vn] (value) {$v^\phi(s_t)$};
        \draw[->] (vn.east) -- (value.west) {};
        
        \node[draw,fill=white] (act) at (action) {$a_t$};
        \draw[-] ($(2.8mm,18mm) + (actions)$) .. controls ($(4mm,0) + (a4.north east)$) and ($(-4mm,0) + (act.north west)$) .. ([yshift=-0.07mm]act.north west) {};
        \draw[-] ($(2.8mm,-2mm) + (actions)$) .. controls ($(4mm,0) + (a0.south east)$) and ($(-4mm,0) + (act.south west)$) .. ([yshift=0.07mm]act.south west) {};
        \draw[->] (act.east) -- ([xshift=8mm]act.east);
    \end{tikzpicture}
    \caption{State $s_t$ is processed by the agent's neural networks. The value network outputs an estimate $v^\phi(s_t)$ of the value function \eqref{eq:value_function}, while the policy network outputs a probability distribution $\pi^\theta(a|s_t)$ on the actions \eqref{eq:policy}. Action $a_t$ is sampled from this probability distribution.}
    \label{fig:rl:agent}
\end{figure}

The PPO algorithm employs two separate neural networks with distinct roles but typically similar structures, see \figref{fig:rl:agent}.
The first is the \emph{policy network}, \idest the actor, which parameterizes the policy $\pi^\theta(a|s)$, explicitly defining the probability distribution over actions given state $s$.
The second is the \emph{value network}, parameterized by parameters $\phi$, which approximates the value function $v^\phi(s)$, estimating the expected cumulative return starting from a state $s$ under the current policy.

Training in PPO relies on iteratively collecting trajectories of state-action pairs from interactions with the environment, then performing updates to the network parameters using gradient ascent on an objective function carefully designed to maintain stability.
The objective used in PPO is based on a clipped surrogate function, formulated to prevent excessively large policy updates.
Specifically, the PPO objective function is defined as:
\begin{equation}
    L^{\text{CLIP}}(\theta) = \mathbb{E}_t \left[\min\left(r_t(\theta)\hat{A}_t,\; \text{clip}(r_t(\theta), 1 - \epsilon, 1 + \epsilon)\hat{A}_t\right)\right],
\end{equation}
where $\hat{A}_t$ is an estimate of the advantage function \eqref{eq:advantage}, computed from the collected trajectories, $r_t(\theta)$ denotes the probability ratio of selecting action $a_t$ under the updated policy relative to the previous policy:
\begin{equation}
    r_t(\theta) = \frac{\pi^\theta(a_t|s_t)}{\pi^{\theta_{\text{old}}}(a_t|s_t)},
\end{equation}
and $\epsilon$ is a hyperparameter controlling the range within which updates to the policy are allowed, ensuring that the policy does not shift excessively during a single update step.

Simultaneously, PPO updates the parameters of the value network by minimizing the squared-error loss between predicted and observed returns, formally expressed as:
\begin{equation}
    L_v(\phi) = \mathbb{E}\left[(v^{\phi}(s_t) - G_t)^2\right],
\end{equation}
where $\phi$ denotes the value network's parameters.
Optionally, an additional entropy term may be included in the combined loss function to encourage exploration and prevent premature convergence to suboptimal deterministic policies.

The PPO update procedure typically involves multiple epochs of mini-batch stochastic gradient ascent on the collected trajectory data, thereby improving sample efficiency and further enhancing stability.
The choice of hyperparameters, including the clipping threshold $\epsilon$, the discount factor $\gamma$, and the learning rate, significantly impacts the performance and convergence behavior of the PPO algorithm.

Due to these desirable properties, PPO has become widely adopted as a powerful baseline for various reinforcement learning applications, ranging from robotic control tasks to complex combinatorial optimization problems, such as the minor embedding scenario addressed in this work \citep{DBLP:conf/icml/SchulmanLAJM15,DBLP:journals/corr/SchulmanWDRK17}.

\subsubsection{Invalid Action Masking}
\label{sec:rl:iam}
In some cases, based on the state the environment is in, some of the actions may not be available. The set of the available actions can be constrained by Invalid Action Masking (IAM) \citep{doi:10.1126/science.aar6404,DBLP:conf/flairs/HuangO22}, ensuring the agent is only able to select valid actions. 
This method consists in masking out invalid actions by setting their probability to zero at the output end of the policy network, thus restricting the set of actions from which the agent chooses.

\section{Reinforcement Learning for Minor Embedding}
\label{sec:rl:mlp}

This section describes our proposed RL model, which is built on simple yet flexible architectures.
The basis of our model is a RL Agent developed using a Multi-Layer Perceptron (MLP) architecture and trained with Proximal Policy Optimization (PPO) to iteratively assign a problem variable to a qubit.
This choice offers several practical advantages: MLPs are easy to implement and relatively fast to train,
while PPO is robust, stable across a wide range of tasks, and exhibits strong empirical performance in high-dimensional action spaces which makes it well-suited for the minor embedding problem.

One disadvantage of the MLP agent architecture is that its structure does not natively allow to leverage graph properties such as permutation invariance, graph symmetries, locality, or connectivity patterns. In order to ensure that those properties are learned by the agent, we design a set of data augmentation strategies to provide the agent with symmetric variants of the same partial minor embedding encouraging more robust and general learning process. While other architectures such as Graph Neural Networks would allow natively to better account for the graph structure, their training poses other challenges and as such we leave them for future work.

\subsection{State Observation and Actions}
\label{sec:rl:state_observation}
In our proposed model the agent is tasked to perform an action based on a partial minor embedding of the problem graph $G$ (with $|G|$ nodes) onto the hardware graph $H$ (with $|H|$ nodes) at each step $t$, where it receives an observation of the state $s_t$. The observation is a one-dimensional array, partitioned into sections, each relating to aspects of either the problem graph $G$ or the hardware graph $H$. Each component corresponds to nodes of the respective graph, following a predetermined mapping. The sate observation has four components: two related to the partial minor embedding and two used to restrict the action space of the agent. The two components related to the partial minor embedding are:
\begin{itemize}
    \item \textbf{Available qubits:}
    This part is denoted by $S_H \in \{0,1\}^{|H|}$, with each component indicating whether that qubit in the hardware graph is available (value 1) or already assigned to a chain and therefore unavailable (value 0).
    At the beginning of each episode, all its values are set to 1.

    \item \textbf{Missing $G$ links:}
    This part is denoted by $S_G \in \mathbb{Z}^{|G|}$, with each component indicating how many problem variables that node still needs to be connected with. More precisely, for every node $G_j$ that is connected to node $G_i$ in the problem graph, the corresponding qubit chain $C_j$ must be connected to $C_i$.
    This measure of how many inter-chain connections are missing serves as an indication to the agent of which parts of the minor embedding are still incomplete.
\end{itemize}

When designing the action space for a RL agent it is important to ensure that the number of actions does not grow excessively with the problem size. A naive space where the agent could be tasked to select both a node from $G$ and a corresponding node from $H$ would result in an action space of size $|G|\times|H|$ which would rapidly become too large.
In order to constrain the dimensionality of the action space to avoid combinatorial growth, we build our RL model so that the agent is provided with a problem variable and has to identify which node of $H$ to allocate it to. At each step $t$ the node from $G$ is selected with a round-robin (RR) strategy that iterates over the problem variables, then the action $a_t$ performed by the agent only selects a node from the hardware graph $H$, which will be added to the chain associated to the current node from $G$. After the action, the next node from $G$ is selected according to the RR sequence, only among those nodes that still lack all the required inter-chain connections.

In order to let the agent understand the RR approach, the last two components of the state observation are:
\begin{itemize}
    \item \textbf{Current node:} 
    Denoted by $S_R \in \{0,1\}^{|G|}$, is a one-hot encoded vector where the node from $G$ which should be added to the minor embedding is marked with $1$.

    \item \textbf{Chain of current node:}
    Denoted by $S_C \in \{0,1\}^{|H|}$, this component is a binary vector indicating which qubits belong to the chain of the current node selected from $G$.
    At the beginning of each episode, all its values are set to 0.
\end{itemize}

Furthermore, the available qubits vector $S_H$ is additionally constrained such that only the qubits that are both available and adjacent to the chain of the currently selected $G$ node in RR are marked as available (value 1).
This results in $S_H$ being the same mask applied via IAM (see \secref{sec:rl:iam}) to the policy, as shown in \figref{fig:rl:iam}.

\begin{figure}[h!]
    \centering
    \begin{tikzpicture}[
        qubit/.style={shape=circle, draw},
        box/.style={inner sep=2mm},
        network/.style={trapezium, trapezium angle=120, minimum width=20mm, minimum height=7.3mm},
    ]
        \coordinate (base) at (0,0);
        
        \node[qubit] (0c) at ($(0,1.5) + (base)$) {0};
        \node[qubit, green, densely dashed, line width=1.5pt] (1c) at ($(0.75,1.5) + (base)$) {1};
        \node[qubit, red, loosely dotted, line width=1.5pt] (2c) at ($(2.25,1.5) + (base)$) {2};
        \node[qubit, brown, dashdotdotted, line width=1.5pt] (3c) at ($(3,1.5) + (base)$) {3};
        \node[qubit] (4c) at ($(1.5,3) + (base)$) {4};
        \node[qubit, blue, line width=1.5pt] (5c) at ($(1.5,2.25) + (base)$) {5};
        \node[qubit] (6c) at ($(1.5,0.75) + (base)$) {6};
        \node[qubit] (7c) at ($(1.5,0) + (base)$) {7};
        
        \path [-] (0c) edge (4c);
        \path [-] (0c) edge (5c);
        \path [-] (0c) edge (6c);
        \path [-] (0c) edge (7c);
        
        \path [-] (1c) edge (4c);
        \path [-, line width=1.5pt] (1c) edge (5c);
        \path [-] (1c) edge (6c);
        \path [-] (1c) edge (7c);
        
        \path [-] (2c) edge (4c);
        \path [-, line width=1.5pt] (2c) edge (5c);
        \path [-] (2c) edge (6c);
        \path [-] (2c) edge (7c);
        
        \path [-] (3c) edge (4c);
        \path [-, line width=1.5pt] (3c) edge (5c);
        \path [-] (3c) edge (6c);
        \path [-] (3c) edge (7c);

        \node[qubit, blue, line width=1.5pt, left=3 of 4c] (0g) {0};
        \node[qubit, red, loosely dotted, line width=1.5pt] (1g) at ($(0g) + (-1,-1)$) {1};
        \node[qubit, green, densely dashed, line width=1.5pt] (2g) at ($(0g) + (1,-1)$) {2};
        \node[qubit, brown, dashdotdotted, line width=1.5pt] (3g) at ($(0g) + (0,-2)$) {3};

        \path[-] (0g) edge (1g);
        \path[-] (0g) edge (2g);
        \path[-] (0g) edge (3g);
        \path[-] (1g) edge (2g);
        \path[-] (1g) edge (3g);
        \path[-] (2g) edge (3g);

        \node[box, right=0.2 of 3g] (gg) {G};
        \node[box, left=0.4 of 7c] (hg) {H};
        \draw[->] (gg) to[bend right] (hg);
        \node[box, left=0.8 of hg] {Minor embedding};

        \node[box, right=3 of 4c, yshift=6mm] (g) {G-related state};

        \matrix (sg) [matrix of nodes, nodes={draw, minimum size=4mm}, nodes in empty cells, column sep=-\pgflinewidth, below=-1mm of g] {0 & 2 & 2 & 2 \\};
        
        \node[left=-1mm of sg] {$S_G$};
        \matrix (sr) [matrix of nodes, nodes={draw, minimum size=4mm}, nodes in empty cells, column sep=-\pgflinewidth, below=-1mm of sg] {0 & 1 & 0 & 0 \\};
        
        \node[left=-1mm of sr] {$S_R$};
        \matrix (idx) [matrix of nodes, white, text=black, nodes={draw, minimum size=4mm}, column sep=-\pgflinewidth, below=-1mm of sr] {0 & 1 & 2 & 3 \\};
        
        \node[box, below=0.2 of idx] (h) {H-related state};
        
        \matrix (sh) [matrix of nodes, nodes={draw, minimum size=4mm}, nodes in empty cells, column sep=-\pgflinewidth, below=-1mm of h] {0 & 0 & 0 & 0 & 1 & 0 & 1 & 1 \\};
        
        \node[left=-1mm of sh] {$S_H$};
        \matrix (sc) [matrix of nodes, nodes={draw, minimum size=4mm}, nodes in empty cells, column sep=-\pgflinewidth, below=-1mm of sh] {0 & 0 & 1 & 0 & 0 & 0 & 0 & 0 \\};
        
        \node[left=-1mm of sc] {$S_C$};
        \matrix (idx2) [matrix of nodes, white, text=black, nodes={draw, minimum size=4mm}, column sep=-\pgflinewidth, below=-1mm of sc] {0 & 1 & 2 & 3 & 4 & 5 & 6 & 7 \\};
    \end{tikzpicture}
    \caption{Example of the state's components in an intermediate phase of the RL minor embedding procedure.
    The agent has to embed a G graph with 4 nodes into an H Chimera graph with 8 nodes.
    There are four chains already embedded in H, solid blue (5), dashed green (1), dotted red (2), dashed and dotted brown (3).
    The current \rr G node is 1 ($S_R$). Node 0 has already completed all connections in the minor, while nodes 1, 2 and 3 are still missing 2 links ($S_G$).
    H nodes 4, 6 and 7 are available ($S_H$) to be added and adjacent to the chain of G node 1, composed of only H node 2 ($S_C$).}
    \label{fig:rl:state}
\end{figure}
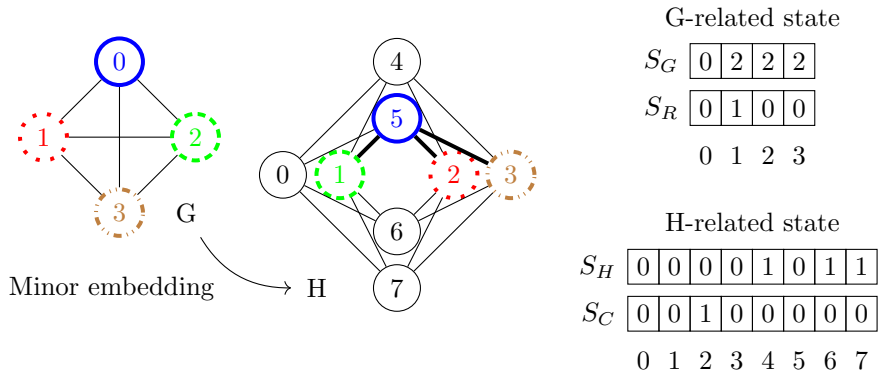

An illustrative example of the entire observation vector during an intermediate minor embedding state is depicted in \figref{fig:rl:state}.

\begin{figure}[h!]
    \centering
    \begin{tikzpicture}[
        qubit/.style={shape=circle, draw},
        box/.style={inner sep=2mm},
        network/.style={trapezium, trapezium angle=120, minimum width=20mm, minimum height=7.3mm},
    ]
        \coordinate (base) at (-7,-3);
        
        \node[qubit, dashed, thick] (0c) at ($(0,1.5) + (base)$) {0};
        \node[qubit, dashed, thick, green] (1c) at ($(0.75,1.5) + (base)$) {1};
        \node[qubit, red, line width=2pt] (2c) at ($(2.25,1.5) + (base)$) {2};
        \node[qubit, dashed, thick] (3c) at ($(3,1.5) + (base)$) {3};
        \node[qubit] (4c) at ($(1.5,3) + (base)$) {4};
        \node[qubit, dashed, thick, blue] (5c) at ($(1.5,2.25) + (base)$) {5};
        \node[qubit] (6c) at ($(1.5,0.75) + (base)$) {6};
        \node[qubit] (7c) at ($(1.5,0) + (base)$) {7};
        
        \path [-] (0c) edge (4c);
        \path [-] (0c) edge (5c);
        \path [-] (0c) edge (6c);
        \path [-] (0c) edge (7c);
        
        \path [-] (1c) edge (4c);
        \path [-, line width=1.5pt] (1c) edge (5c);
        \path [-] (1c) edge (6c);
        \path [-] (1c) edge (7c);
        
        \path [-] (2c) edge (4c);
        \path [-, line width=1.5pt] (2c) edge (5c);
        \path [-] (2c) edge (6c);
        \path [-] (2c) edge (7c);
        
        \path [-] (3c) edge (4c);
        \path [-] (3c) edge (5c);
        \path [-] (3c) edge (6c);
        \path [-] (3c) edge (7c);

        \node[fit=(0c)(1c)(2c)(3c)(4c)(5c)(6c)(7c)] (chi) {};
        
        \node[box, draw] (st) {Input State};
        \node[network, draw, below of=st] (pn) {Policy Network};
        
        \matrix (out) [matrix of nodes, nodes={draw, minimum size=4mm}, nodes in empty cells, column sep=-\pgflinewidth, below of=pn] {
             &  &  &  &  &  &  & \\};
        \begin{scope}[on background layer]
        \foreach [count=\i from 1] \shade in {21, 4, 13, 7, 18, 2, 6, 32}
            \fill[blue!\shade] (out-1-\i.south west) rectangle (out-1-\i.north east);
        \end{scope}
        
        \matrix (iam) [matrix of nodes, nodes={red, draw, minimum size=4mm}, nodes in empty cells, column sep=-\pgflinewidth, below=-1mm of out] {
            &  &  &  &[0.2mm] |[draw=none]| &  &[0.2mm] |[draw=none]| & |[draw=none]| \\};
        \begin{scope}[on background layer]
        \foreach \i in {1, 2, 3, 4, 6}
            \fill[pattern={Dots[angle=45,distance=1.5pt]}, pattern color=red] (iam-1-\i.south west) rectangle (iam-1-\i.north east);
        \end{scope}
        
        \matrix (nout) [matrix of nodes, nodes={draw, minimum size=4mm}, nodes in empty cells, column sep=-\pgflinewidth, below=-1mm of iam] {
             &  &  &  &  &  &  & \\};
        \begin{scope}[on background layer]
        \foreach \i/\shade in {5/32, 7/10, 8/58}
            \fill[blue!\shade] (nout-1-\i.south west) rectangle (nout-1-\i.north east);
        \end{scope}

        \matrix (idx) [matrix of nodes, nodes={draw, white, text=lightgray, minimum size=4mm}, column sep=-\pgflinewidth, below=-2mm of nout] {
             0 & 1 & 2 & 3 & |[text=black]| 4 & 5 & |[text=black]| 6 & |[text=black]| 7 \\};

        \node[right=1mm of out] (tout) {Original output};
        \node[right=1mm of iam] (tiam) {Invalid Action Masking};
        \node[right=1mm of nout] (tnout) {Masked probability};

        \node[box, left=0.4 of iam, align=center] (rr) {Round-robin\\chain};

        \node[circle, fill, above=0.4 of 2c, minimum size=3, inner sep=0pt, outer sep=0pt] (pnt) {};

        \draw[->] (st.south) -- (pn.north);
        \draw[->] (pn.south) -- (out.north);
        \draw[->, red, thick] (2c.south) .. controls ($(2c) + (0,-1)$) and ($(rr) + (-1.4,0)$) .. (rr.west) {};
        \draw[->, red] (rr.east) -- (iam.west) {};
        \path[->, dashed] (pnt) edge[bend left=12] (st.west);

    \end{tikzpicture}
    \caption{Application of Invalid Action Masking to the agent's policy.
    On the left there is the partial minor embedding on which the agent is working, with three different chains (green, red and blue) containing respectively nodes 1, 2 and 5.
    This is the environment's state, whose observation is received by the agent on the right.
    The policy network gets the state as input and outputs a series of values, one for each possible action, \idest nodes from 0 to 7.
    Since the current \rr G node is the one corresponding to the red chain (containing only H node 2), the applied mask comprises all the nodes in the chain (2), the ones that are not adjacent to the chain (0, 1, 3) and the ones already embedded in other chains (1, 5).
    Therefore, the only nodes which will have a non-null value in the actual output will be 4, 6 and 7.}
    \label{fig:rl:iam}
\end{figure}
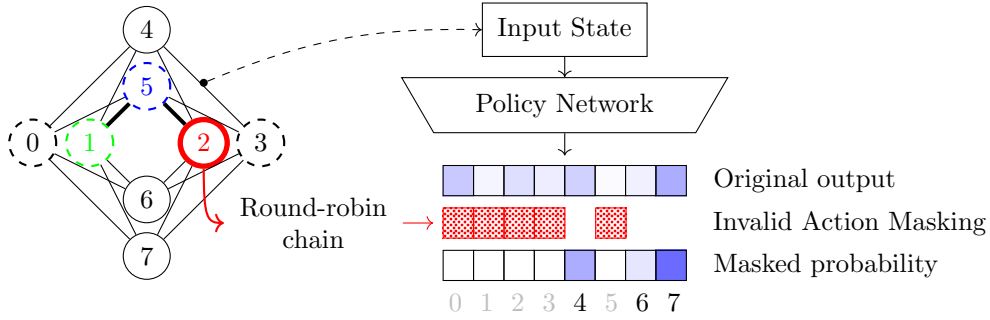

\subsection{Reward Function}
\label{sec:rl:reward}
After selecting an action based on the current state, the agent receives a scalar reward from the environment. The design of the reward function depends directly on the objective one wants to optimize during minor embedding. Given that our main goal is to obtain a valid minor but also to minimize the length of the chains, each action gives a fixed negative reward, so that the agent is encouraged to maximize the cumulative reward by creating shorter chains.
In particular, in the experiments presented in \secref{sec:rl:results}, we use a negative reward of $-0.1$ for each action.

At the same time, one of the advantages of the RL framework is its flexibility, which allows the reward function to be changed or redesigned depending on different optimization goals. For example, it would be possible to define a reward based on the actual solution quality of the embedded problem when executed on a quantum annealer. Another option would be to use a sparse reward scheme, where the agent receives a non-zero reward only when a complete and valid minor is produced. While these examples show the adaptability of the reward mechanisms, they would each present new challenges to address and, therefore, the exploration of these options constitutes a broad research direction that goes beyond the scope of this paper.

\section{Experimental Protocol}
The goal of the experimental analysis is to assess the effectiveness of the proposed RL model in two scenarios: the minor embedding of fully connected problem graphs, and that of randomly generated ones.
In the fully connected scenario, the agent is trained to perform \ac{ME} of a specific fully connected graph G onto a specific topology graph H.
In the random graph scenario, each agent is trained to perform \ac{ME} of problem graphs G with different sizes and densities onto a specific topology graph H, with the objective of evaluating whether the model is capable of generalizing to unseen instances and varied topologies.

While the internal topology of graph $G$ differs, both scenarios share the same graph sizes. The number of nodes of $G$ ranges from 3 to 10. The process used to generate the random graphs is described in \secref{sec:rl4me:dataset}.
For what concerns the hardware graph H, we generate them based on the number of unit cells per side, $H_{size}$, with two topologies: Chimera, with a number of nodes ranging from 32 to 2048 ($H_{size} \in [2,16]$), and Zephyr, with a number of nodes ranging from 160 to 2176 ($H_{size} \in [2,8]$).
These topologies correspond to the  oldest (Chimera) and newest (Zephyr) topology available at the time these experiments were conducted, and their $H_{size}$ has been chosen to ensure the number of qubits is comparable.

\subsection{Randomly Generated Graphs}
\label{sec:rl4me:dataset}
In order to evaluate the effectiveness of our proposed RL model we train and evaluate it also on a set of randomly generated problem graphs\footnote{The dataset of randomly generated graphs is available at \url{https://github.com/qcpolimi/RLxME_Dataset}}. While it could be argued that a random graph is not necessarily representative of the types of topologies and structures one could observe when applying ME to real problems, these random graphs serve the purpose of testing the RL model on a set of highly heterogeneous problems.

We generate graphs G ranging in size from 3 nodes upwards, and with varying number of edges. For a graph with $n = |G|$ nodes, the number of edges ranges from $n-1$ to $\frac{n (n - 1)}{2}$.
We also ensure that all graphs are connected, \idest from every node there should exist a path to each other node in the graph. For each number of nodes we aim to generate 1000 graphs to use during the training phase and an additional 250 to use for testing. Based on the experimental results (see \secref{sec:rl:results}) the largest problem graph tested is $n=10$.

The process used to generate the graphs depends on their number of nodes:
\begin{itemize}
    \item \textbf{Graphs with 3 to 5 nodes:} Starting from a fully connected graph, we generate a set of all the graphs that can be obtained from it by removing one edge. Then, for each of those graphs we repeat the process removing one further edge. We discard all graphs that have non-connected components. Due to the small number of nodes it is not possible to reach the desired number of graphs, as such the same training graph will be sampled multiple times during training.
    \item \textbf{Graphs with 6 to 8 nodes:} We follow the same approach used for the smaller graphs with 3 to 5 nodes, but we also ensure to keep only one instance of graphs that are \emph{isomorphic}.\footnote{Verifying whether two graphs are isomorphic is computationally expensive. We rely on an approximate method of the \texttt{networkx} package which checks the degree sequence of the nodes and identifies rapidly graphs that are surely non-isomorphic based on that \url{https://networkx.org/documentation/stable/reference/algorithms/generated/networkx.algorithms.isomorphism.faster_could_be_isomorphic.html}} Once this phase is completed, we obtain a set of graphs each defining an isomorphism class. Based on their number we compute how many instances for each class are required in order to reach the target number of training and testing graphs, and finally we generate them by applying a random node permutation. We always ensure that no two graphs are identical.
    \item \textbf{Graphs with more than 9 nodes:} For graphs of this size the exhaustive generation of both instances as well as one instance for each isomorphism class becomes unfeasible. The approach we adopt here is to first identify what is the minimum and maximum number of edges these graphs could have (\idest $n-1$ to $\frac{n (n - 1)}{2}$), then we compute how many graph instances for each of the values in this range we should generate in order to reach the target number of training and testing graphs. Given a number of edges, a new graph instance is generated at random.\footnote{We rely on the following function of the \texttt{networkx} package \url{https://networkx.org/documentation/stable/reference/generated/networkx.generators.random_graphs.gnm_random_graph.html}} We guarantee that the set does not contain isomorphic graphs.
\end{itemize}

The testing data is created with a 20\% random holdout of the graphs, stratified on the number of nodes, hence 1000 are used for training and 250 for testing for each node size. The process accounts for the strategies adopted to construct graphs of each size: for small graphs from 3 to 5 nodes we apply a simple random holdout; for graphs of 6 to 8 nodes the random holdout is stratified on the isomorphism classes; finally, for graphs of more than 9 nodes the holdout is stratified on the number of edges in the graph.
Note that while the testing data will never contain graphs that also appear in the training data, the testing data of small graphs (from 3 to 5 nodes) may contain isomorphic ones. We believe this is not a problem as it serves as a way to test whether the RL model has learned to model graph symmetries, and it only occurs for small graphs.

\subsection{Training Protocol}
The agent is trained using \acl{PPO} (see \secref{sec:rl:ppo}) and \acl{IAM} (see \secref{sec:rl:iam}, an example shown in \figref{fig:rl:iam}).

In the fully connected graph scenario the agent is trained on performing minor embedding of a specific graph G on a specific topology H, therefore it only observes a single fully connected graph. The training budget is of 1 million steps, however, as we observed in the results (see \secref{sec:training_convergence}), the agent tends to reach convergence \textbf{much faster} especially when $H$ is small. This means that the main \ac{RL} interaction loop described in \secref{sec:bg:rl} is repeated that number of times before stopping the training phase.

In the random graph scenario the training is performed by feeding all the training graphs to the agent ordered by increasing number of nodes. The purpose of this strategy is to allow the agent to initially focus on structurally simpler instances.  Given that the training process is performed on many more graphs and to esure the agent has the time to train on each of them, the budget in this experiment is of 3 million steps. 

During training it is guaranteed that at least $10^3$ graphs for each number of nodes are used. When the number of nodes is small, \idest 3-5, the number of existing training graphs is lower than this threshold, therefore the same training graph may be sampled multiple times. Once all graphs up to $|G|=10$ have been sampled, the process repeats from $|G|=3$, until the step budget is depleted.

In both scenarios the training is done on 10 independent agents each starting from a different random initialization. 
During testing, each agent is tasked to generate 10 minor embeddings for each test graph. In the fully connected scenario the testing graph is the same the agent has been trained on, while in the random graph scenario the testing graphs are those held-out as described in \secref{sec:rl4me:dataset}.

\subsection{Training Graph Augmentations}
\label{sec:model:da}
The hardware graphs $H$ possess several \emph{symmetries}, stemming from their graph nature but also from the highly regular structure of the hardware topologies. These include global symmetries such as node permutations, reflections, and rotations of the layout of the hardware graph, particularly when $H$ is formed by repeating patterns like Zephyr or Chimera unit cells. As a consequence, multiple minors can be functionally equivalent up to a relabelling or spatial transformation of the physical qubits. For example, a valid minor can be rotated $90^\circ$ across a regular 2D layout of cells or flipped horizontally without altering its correctness or quality. This permutation invariance implies that an optimal policy for minor embedding should ideally be invariant (or at least equivariant) under these transformations. 

However, MLP-based policy architectures lack any inherent mechanism to model this invariance. Since they operate on a flattened, fixed-size observation vector, the agent will consider two isomorphic states as completely different, leading to slow training and poor generalization. In order to improve the robustness of the MLP agent to those symmetries, at each step of the training process we can modify the partial minor embedding by applying an isomorphic transformation on the environment. This strategy is inspired by similar techniques in image recognition and structured decision-making problems \citep{DBLP:journals/jbd/ShortenK19, Silver2016}. The core idea is to expose the agent to different but semantically equivalent versions of the environment state, so that the learned policy becomes more robust to the symmetries of the underlying problem.

\begin{figure}
    \centering
    \hfill
    \begin{subfigure}[b]{0.3\textwidth}
        \centering
        \includegraphics[width=\textwidth]{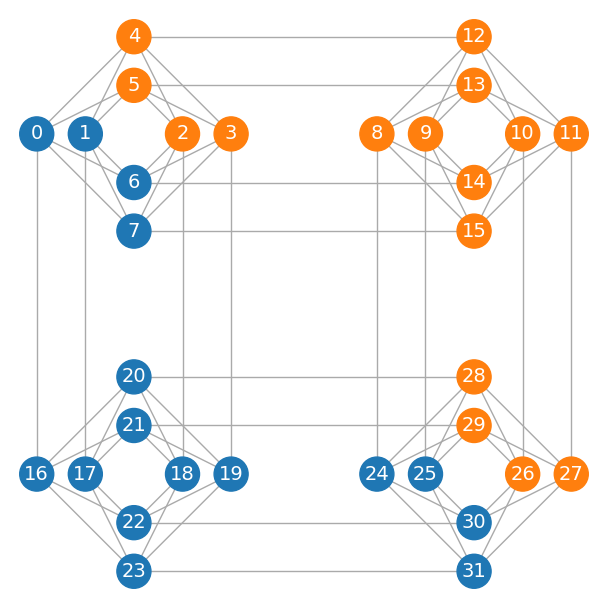}
        \caption{Chimera graph with $H_{size} = 2$}
        \label{fig:augmentations:original}
    \end{subfigure}
    \hfill
    \begin{subfigure}[b]{0.3\textwidth}
        \centering
        \includegraphics[width=\textwidth]{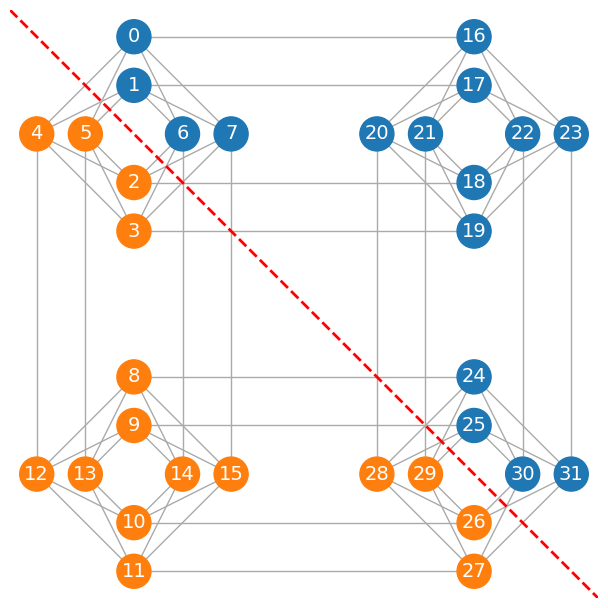}
        \caption{Symmetry applied on the main diagonal}
        \label{fig:augmentations:symmetry}
    \end{subfigure}
    \hfill
    \begin{subfigure}[b]{0.3\textwidth}
        \centering
        \includegraphics[width=\textwidth]{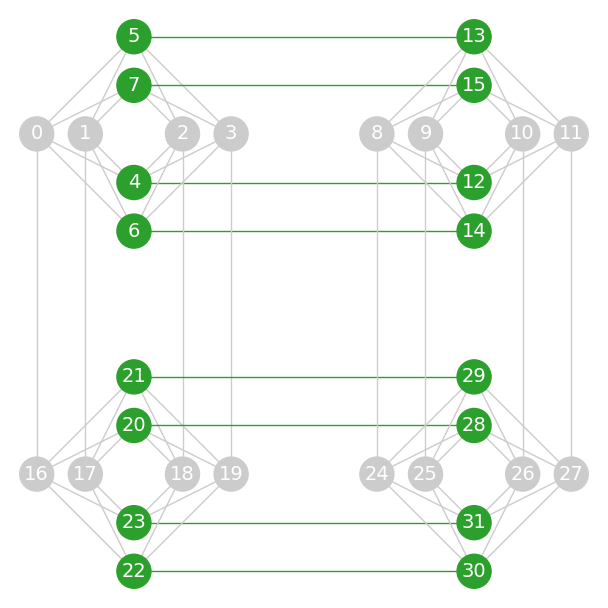}
        \caption{Permutation applied on vertical nodes}
        \label{fig:augmentations:permutation}
    \end{subfigure}
    \caption{Example of augmentations on a Chimera hardware graph with $H_{size} = 2$ (4 unit cells). From the original graph (\ref{fig:augmentations:original}) we show the main diagonal symmetry (\ref{fig:augmentations:symmetry}) and the permutation on vertical qubits (\ref{fig:augmentations:permutation}). Colours are used just to improve the visibility of the augmentations.}
    \label{fig:augmentations}
\end{figure}

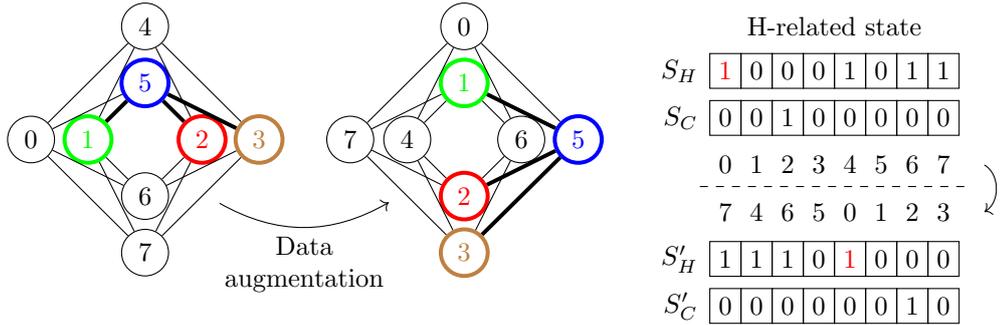
\begin{figure}
    \centering
    \begin{tikzpicture}[
        qubit/.style={shape=circle, draw},
        box/.style={inner sep=2mm},
        network/.style={trapezium, trapezium angle=120, minimum width=20mm, minimum height=7.3mm},
    ]
        \coordinate (base) at (-7,-3);
        \coordinate (base) at (0,0);
        \coordinate (transform) at ($(base) + (4.2, 0)$);
        
        \node[qubit] (0c) at ($(0,1.5) + (base)$) {0};
        \node[qubit, green, line width=1.5pt] (1c) at ($(0.75,1.5) + (base)$) {1};
        \node[qubit, red, line width=1.5pt] (2c) at ($(2.25,1.5) + (base)$) {2};
        \node[qubit, brown, line width=1.5pt] (3c) at ($(3,1.5) + (base)$) {3};
        \node[qubit] (4c) at ($(1.5,3) + (base)$) {4};
        \node[qubit, blue, line width=1.5pt] (5c) at ($(1.5,2.25) + (base)$) {5};
        \node[qubit] (6c) at ($(1.5,0.75) + (base)$) {6};
        \node[qubit] (7c) at ($(1.5,0) + (base)$) {7};
        
        \path [-] (0c) edge (4c);
        \path [-] (0c) edge (5c);
        \path [-] (0c) edge (6c);
        \path [-] (0c) edge (7c);
        
        \path [-] (1c) edge (4c);
        \path [-, line width=1.5pt] (1c) edge (5c);
        \path [-] (1c) edge (6c);
        \path [-] (1c) edge (7c);
        
        \path [-] (2c) edge (4c);
        \path [-, line width=1.5pt] (2c) edge (5c);
        \path [-] (2c) edge (6c);
        \path [-] (2c) edge (7c);
        
        \path [-] (3c) edge (4c);
        \path [-, line width=1.5pt] (3c) edge (5c);
        \path [-] (3c) edge (6c);
        \path [-] (3c) edge (7c);
        
        \node[qubit] (7t) at ($(0,1.5) + (transform)$) {7};
        \node[qubit] (4t) at ($(0.75,1.5) + (transform)$) {4};
        \node[qubit] (6t) at ($(2.25,1.5) + (transform)$) {6};
        \node[qubit, blue, line width=1.5pt] (5t) at ($(3,1.5) + (transform)$) {5};
        \node[qubit] (0t) at ($(1.5,3) + (transform)$) {0};
        \node[qubit, green, line width=1.5pt] (1t) at ($(1.5,2.25) + (transform)$) {1};
        \node[qubit, red, line width=1.5pt] (2t) at ($(1.5,0.75) + (transform)$) {2};
        \node[qubit, brown, line width=1.5pt] (3t) at ($(1.5,0) + (transform)$) {3};
        
        \path [-] (0t) edge (4t);
        \path [-] (0t) edge (5t);
        \path [-] (0t) edge (6t);
        \path [-] (0t) edge (7t);
        
        \path [-] (1t) edge (4t);
        \path [-, line width=1.5pt] (1t) edge (5t);
        \path [-] (1t) edge (6t);
        \path [-] (1t) edge (7t);
        
        \path [-] (2t) edge (4t);
        \path [-, line width=1.5pt] (2t) edge (5t);
        \path [-] (2t) edge (6t);
        \path [-] (2t) edge (7t);
        
        \path [-] (3t) edge (4t);
        \path [-, line width=1.5pt] (3t) edge (5t);
        \path [-] (3t) edge (6t);
        \path [-] (3t) edge (7t);

        \node[box, below left=0.4 of 3c] (as) {};
        \node[box, below right=0.4 of 7t] (ae) {};
        \draw[->] (as) to[bend right] node[midway, below, align=center] {Data\\augmentation} (ae);

        \node[box, right=3.2 of 0t] (h) {H-related state};
        
        \matrix (sh) [matrix of nodes, nodes={draw, minimum size=4mm}, nodes in empty cells, column sep=-\pgflinewidth, below=-1mm of h] {|[text=red]| 1 & 0 & 0 & 0 & 1 & 0 & 1 & 1 \\};
        \node[left=-1mm of sh] {$S_H$};
        
        \matrix (sc) [matrix of nodes, nodes={draw, minimum size=4mm}, nodes in empty cells, column sep=-\pgflinewidth, below=-1mm of sh] {0 & 0 & 1 & 0 & 0 & 0 & 0 & 0 \\};
        \node[left=-1mm of sc] {$S_C$};
        
        \matrix (idx) [matrix of nodes, white, text=black, nodes={draw, minimum size=4mm}, column sep=-\pgflinewidth, below=-1mm of sc] {0 & 1 & 2 & 3 & 4 & 5 & 6 & 7 \\};

        \draw[-, dashed] ($(idx.south west) + (0,0.075)$) to ($(idx.south east) + (0,0.075)$);
        
        \matrix (tidx) [matrix of nodes, white, text=black, nodes={draw, minimum size=4mm}, column sep=-\pgflinewidth, below=-1mm of idx] {7 & 4 & 6 & 5 & 0 & 1 & 2 & 3 \\};
        
        \matrix (tsh) [matrix of nodes, nodes={draw, minimum size=4mm}, nodes in empty cells, column sep=-\pgflinewidth, below=-1mm of tidx] {1 & 1 & 1 & 0 & |[text=red]| 1 & 0 & 0 & 0 \\};
        \node[left=-1mm of tsh] {$S^\prime_H$};
        
        \matrix (tsc) [matrix of nodes, nodes={draw, minimum size=4mm}, nodes in empty cells, column sep=-\pgflinewidth, below=-1mm of tsh] {0 & 0 & 0 & 0 & 0 & 0 & 1 & 0 \\};
        \node[left=-1mm of tsc] {$S^\prime_C$};

        \draw[->] ($(idx.east) + (0.2, 0)$) to[bend left=60] ($(tidx.east) + (0.2, 0)$);
    \end{tikzpicture}
    \caption{Example of data augmentation on the same state shown in \figref{fig:rl:state}.
    A permutation on the horizontally-place nodes is applied after a clockwise rotation of $90^\circ$.
    On the right the effect of such transformation is shown on the part of the state related to H.
    The original values (above) are reordered accordingly to the transformation, as if they occupy the new place assigned to them (7 instead of 0, 4 instead of 1 and so on).}
    \label{fig:rl:da}
\end{figure}

The data augmentation strategies we implemented can be grouped in three categories:
\begin{itemize}
    \item \textbf{90° Rotations:} clockwise and counter-clockwise.
    \item \textbf{Mirroring:} along the vertical axis, horizontal axis, main diagonal (see \figref{fig:augmentations:symmetry}) and anti-diagonal.
    \item \textbf{Permutation:} along the vertical axis (given a row of the topology we apply the same permutation to all vertical qubits, with a different permutation applied on each row of the topology, see \figref{fig:augmentations:permutation}), similarly along the horizontal axis.
\end{itemize}
This corresponds to a total of 8 possible data augmentations.
At each training step, a set of zero or more augmentations, without repetitions, is selected at random and applied to the $H$ graph.
Each transformation results in a shuffled version of the state: if, for example, nodes 0 and 7 are swapped, the state vector is also updated accordingly. This process is illustrated in \figref{fig:rl:da}.

This strategy is conceptually aligned with successful practices in other domains, such as data augmentation in computer vision, where image transformations (e.g., flips, crops, rotations) help convolutional networks learn translational invariance. In reinforcement learning, analogous techniques have been employed to improve sample efficiency and policy generalization, especially in spatially structured environments such as robotic control or grid-based navigation tasks.
In our case, the augmentation serves to regularize the policy and promote invariance to graph isomorphisms and spatial symmetries.

While effective in the small to medium-scale setting, this data augmentation approach faces inherent limitations as the size of the hardware graph $H$ increases. The number of possible augmentations grows rapidly, and the diversity of states the agent needs to generalize over increases combinatorially. Moreover, designing augmentation rules that remain valid and computationally tractable on large, irregular or defected hardware topologies becomes increasingly challenging. Thus, although data augmentation can improve the robustness of the MLP-based agent, especially on randomly generated graphs, it cannot fully resolve the lack of topology awareness in the architecture. While there are architectures that natively model this, such as Graph Neural Networks (GNNs), we leave their exploration as a future work.

\subsection{Environment Initialization}
Each time the reinforcement learning environment completes an episode, its internal state, described in detail in \secref{sec:rl:state_observation}, is re-initialized. This is necessary to ensure that the environment begins from a clean and valid configuration. In particular, the mask of the available qubits $S_H$ is reset, any previous minor assignments are cleared and the round-robin pointer is set to the first node of $G$.

\subsection{Implementation Details}
The reinforcement learning agent is trained using the \texttt{stable-baselines3} library \citep{stable-baselines3}, which provides a robust implementation of the Proximal Policy Optimization (PPO) algorithm.
The environment is built upon the \texttt{gymnasium} \citep{towers2024gymnasium} interface and simulates the minor embedding process in order to prepare the state observation vector (see \secref{sec:rl:state_observation}).
The hyperparameters used for PPO are the default ones from \texttt{stable-baselines3}.

\subsection{Evaluation Metrics}
\label{sec:MLP-metrics}
The effectiveness of our RL model is evaluated using two main criteria: \emph{Success Rate} and \emph{Qubit Efficiency Ratio}, to assess its ability to generate valid minors as well as its efficiency in using the available qubits.

\subsubsection*{Success Rate (SR)}
The \emph{Success Rate (SR)} measures the quota of valid minors over those generated by the 10 RL agents trained on different random intializations. For the fully connected scenario, the models are tested only on 1 testing graph, for the random scenario the model is tested on 250 graphs for each number of nodes. For each testing graph, each of the 10 RL agents generate 10 minors each.

\subsubsection*{Qubit Efficiency Ratio (QER)}
We also assess the quality of the minor embeddings by comparing the number of physical qubits used by the model against those used by the \texttt{minorminer} baseline. This is captured by the \emph{Qubit Efficiency Ratio (QER)}, defined as:
\begin{equation}
\text{Qubit Efficiency Ratio} = \frac{|\mathcal{E}_{\text{MM}}|}{|\mathcal{E}_{\text{RL}}|},
\label{eq:qur}
\end{equation}
where $|\mathcal{E}_{\text{RL}}|$ denotes the number of qubits used by the model in its best minor embedding (\idest the one using the fewest qubits among the different testing trials), and $|\mathcal{E}_{\text{MM}}|$ is the minimum number of qubits used by \texttt{minorminer} on the same problem graph, among 100 minors for the fully connected scenario and 10 minors for each of the testing graphs in the randomly generated graphs scenario.
Given that \texttt{minorminer} typically produces highly compact minors for problems of the scale used in this work, it serves as a near-optimal baseline for this comparison.
A higher QER value (\idest closer to 1) indicates that the model's minor embedding is close in quality to the baseline.
In contrast, excessive qubit usage \wrt to \texttt{minorminer} will increase the denominator in Eq.~\eqref{eq:qur}, pushing the ratio towards 0.

\section{Results and Discussion}
\label{sec:rl:results}

In this section we present the results of the experimental analysis along the two scenarios: fully connected graphs and randomly generated ones. For the first scenario we use both topologies Chimera and Zephyr and discuss in details the results on each of them. For the random graph scenario we instead focus on Zephyr.

\subsection{Fully Connected Problem Graphs}

\subsubsection{Chimera Topology}
Chimera is the oldest topology where each qubit is connected only to up to 6 other ones, as such it represents a hardware graph where the minor embedding will tend to be rather large. A selection of the results for the Chimera topology is reported in \tabref{table:rl:mlp:chimera}, the full results are reported in Appendix~\ref{app:results} (Tables~\ref{table:rl:mlp:chimera:g3-6} and \ref{table:rl:mlp:chimera:g7-10}).

\begin{table}
    \centering
    \caption{Success rate, as well as average and standard deviation of the number of qubits required by the minor embedding built by the RL agent on fully connected problem graphs G mapped to Chimera topology graphs H. This table shows a slice of the full results, which are reported in Appendix~\ref{app:results} (Tables~\ref{table:rl:mlp:chimera:g3-6} and \ref{table:rl:mlp:chimera:g7-10}).}
    \label{table:rl:mlp:chimera}
    \footnotesize
    \input{mlp_chimera}
\end{table}

\begin{figure}
    \centering
    \includegraphics[width=0.5\linewidth]{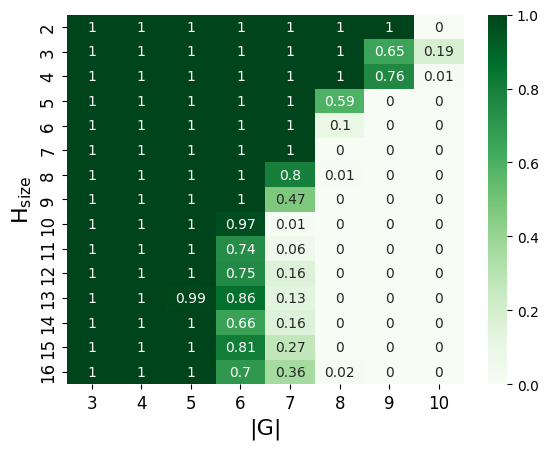}
    \caption{Success rate for the fully connected scenario on Chimera.}
    \label{fig:rl:mlp:success}
\end{figure}

\paragraph{Success Rate}
First, we will discuss the results of the base version of the agent, without data augmentations.
By analysing the success rate of the mionor embedding \idest the quota of minor embeddings generated by the agent that constitute a correct and complete minor of graph $G$ on the hardware graph $H$ we can see how for small $H_{size}$ (2 to 6) the success rate is very high for $|G| \leq 8$. As $H_{size}$ increases we can see that the success rate drops sharply for larger $G$, to the point of failing to produce a valid minor at $H_{size}=|G|=8$. Instead, for smaller $G$ the success rate remains rather stable, for example, $|G|=6$ has a success rate above $90\%$ for $H_{size}\leq 10$ and then a success rate oscillating between  $60\%$ and $75\%$ as $H$ grows larger up to 16.
These results indicate that the success rate is much more sensitive to the size of $G$ than it is to the size of $H$. This effect is immediately visible from \figref{fig:rl:mlp:success}. The reasons for this can be several, but it is apparent that, as $G$ becomes larger, the agent struggles to model the increased complexity of the minor embedding process with many increasingly long chains that need to be connected across several cells. Which points to the need to strengthen the ability of the agent to model complex graph structures.

The data augmentation aims to tackle this limitation. Since the agent is provided with graphs and minors that have been permuted, in different ways, the aim is that this should improve the ability of the agent to identify the right correlations in the data and learn a more robust latent representation of the overall minor structure. If we look at the results in \tabref{table:rl:mlp:chimera} we can compare the success rate of the base model with that where the data augmentation is applied. The results are mixed. If we go back to the previous example of $|G|=6$, the success rate of the base model is above $90\%$ for $H_{size}\leq 10$ but that of the data augmentation version begins to drop earlier, with $H_{size}=10$ having a success rate of $77\%$. For larger $H$ the success rate oscillates, as in the base model, but not consistently. For example, for $H_{size}=14$ the base version has a success rate of $66\%$ while the data augmentation version of $69\%$. On the opposite end is the immediately following $H_{size}=16$ where the base version has a success rate of $70\%$ while the data augmentation version of $50\%$, a 20 points difference. While this seems to suggest the data augmentation is not beneficial, in this experiment at least, as $G$ becomes larger the conclusions change. If we look at $|G|=7$ (see \tabref{table:rl:mlp:chimera:g7-10}) the base version of the agent starts to struggle with $H_{size}=8$ and becomes largely ineffective for larger $H$ oscillating between a success rate of $1\%$ and $36\%$. However, it is here that we see the benefit of the data augmentation which is able to keep the success rate higher for intermediate values of $H_{size}$. For example, for $H_{size}=10$ the base version has a success rate of $1\%$ while the data augmentation version of $32\%$, or for $H_{size}=12$ where the base version has a success rate of $16\%$ while the data augmentation version of $30\%$. However, these gains do not remain present at even higher $H$. For the largest $G$, the data augmentation is sometimes the only version that is able to provide any minor embedding at all, in particular for $H_{size}=7,9,11,14$ where the base version always fails.
Overall, these results indicate that the data augmentation is not a consistently better strategy, and point to the possibility of treating it as a hyperparameter to be activated or not based on the results of the model.

\paragraph{Number of Qubits}
Another important dimension to consider is how many qubits are used by a successful minor embedding.
Clearly, the first aim is to obtain a valid minor, however once that has been achieved, smaller minor embeddings are preferable.
This is due to a number of factors but mainly to how larger minor embeddings require longer chains of connected qubits.
In practice, the physical consequence of a long chain is that it is going to be more difficult for the qubits to change their value as doing so requires to overcome an energy barrier that does not only depend on the other chains the qubit is connected to, but also on all the other qubits of the same chain that are strongly coupled to each other.
\tabref{table:rl:mlp:chimera} shows the average and standard deviation of the number of qubits required by the successful minor embeddings produced by the RL agent.
If we focus on small $H_{size}$ we can see how the number of qubits does not change significantly with increasing sizes of $G$.
For example, when $H_{size}=4$ a graph $|G|=4$ with the base agent requires 6 qubits, while a graph $|G|=6$ requires 14 and finally $|G|=8$ requires 29.
This indicates that the agent is rather efficient in this setting.
If we look at increasing sizes of $H$, however, we can see a marked increase in the number of required qubits, which surpasses 1000.
The first question to ask is why should it be necessary to use a higher number of qubits to embed the same graph on a larger hardware graph.
The answer is that it should not.
This effect shows that the agent is struggling to correctly model and explore this larger hardware graph.
The limited modelling capacity of the MLP-based agent is also compounded by the known difficulties of RL on large action spaces which would begin to play a role for large $H$ graphs.
Fortunately, this limitation can be easily mitigated by first attempting to embed the graph on a smaller $H$ and then move to larger ones only when needed.

If we compare the effect of the data augmentation, again we can see that no consistent pattern emerges, with a few exceptions.
For $|G|=4$ the number of qubits required by the augmented versions is consistently higher than the base version, sometimes by a considerable margin.
Consider for example $H_{size}=16$, if we look at $|G|=4$ we can see how the base version requires only an average of 167 qubits, while the data augmentation version requires 642.
If we move to larger $G$ this difference disappears.
When $H_{size}=16$ and $|G|=6$ the base version requires approximately 1300 qubits while the data augmentation one approximately 1450.
Then, the difference becomes smaller and smaller as $G$ becomes larger.
The reason for this effect may be explained by how for large $H$ and relatively small $G$ the base version of the agent finds easier to use one cell or its immediate surrounding ones, while the augmented ones are pushed to explore the hardware graphs much more, hitting into the limited modelling capacity of this architecture and resulting in a successful, but inefficient, minor embedding.

\begin{figure}
    \centering
    \includegraphics[width=0.5\linewidth]{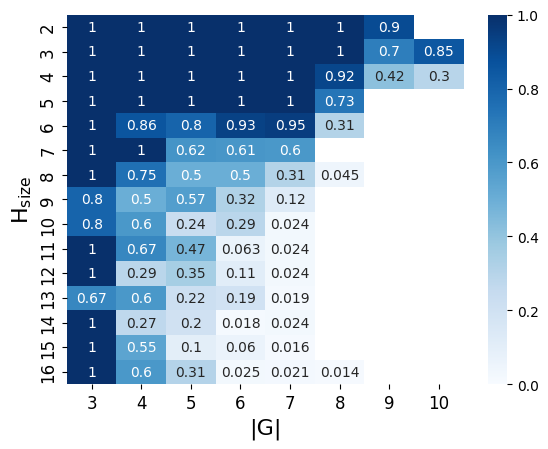}
    \caption{Qubit Efficiency Ratio for the fully connected scenario on Chimera.}
    \label{fig:rl:mlp:chimera_qur}
\end{figure}

\figref{fig:rl:mlp:chimera_qur} shows a visualization of the Qubit Efficiency Ratio, the ratio between the number of qubits required by the minor embeddings produced by \texttt{minorminer} versus the ones identified by the agent without augmentations.
We can see how for $|G|\leq 7$ and for small $H$ graphs (up to 200 qubits) the agent uses as many qubits as \texttt{minorminer}.
This is typically due to the fact that the minor embedding is either optimal or very close to optimal.
As the size of $G$ increases we can see that the minor embedding becomes less efficient and the agent finds minors requiring from slightly more to up to three times more qubits.
A much more marked effect is present by increasing the size of $H$ while keeping $G$ smaller.
As we discussed previously, this is a region where the agent is less efficient, with the augmented versions requiring a very large number of qubits compared to the base version.
The base version itself however already requires many more qubits than \texttt{minorminer}.
For example, for $H_{size}=12$ and $|G|=6$ the agent requires ten times as many qubits as \texttt{minorminer}, and almost fifty times as many when $|G|=7$.
This again points to how the modelling capability of the agent and the large action space make this a challenging scenario.

\begin{table}
\centering
\caption{Success rate, as well as average and standard deviation of the number of qubits required by the minor embedding built by the RL agent on fully connected problem graphs G mapped to Zephyr topology graphs H. This table shows a slice of the full results, which are reported in Appendix~\ref{app:results} (Tables~\ref{table:rl:mlp:zephyr:g3-6} and \ref{table:rl:mlp:zephyr:g7-10}).}
\label{table:rl:mlp:zephyr}
\footnotesize
\input{mlp_zephyr}
\end{table}

\subsubsection{Zephyr Topology}
Zephyr is the newest topology where each qubit is connected to up to 20 other ones, as such it represents a hardware graph where the minor embedding will tend to be rather compact \wrt Chimera.
A selection of the results for the Zephyr topology are reported in \tabref{table:rl:mlp:zephyr}, the full results are reported in Appendix~\ref{app:results} (Tables~\ref{table:rl:mlp:zephyr:g3-6} and \ref{table:rl:mlp:zephyr:g7-10}).

\paragraph{Success Rate} The first observation we can draw is that all experiments exhibit a success rate of $100\%$ both increasing the size of $G$ and by increasing the size of $H$. It should also be noted that the size of cells in Zephyr is larger than in Chimera, so $H_{size}=8$ for Zephyr is comparable to $H_{size}=16$ for Chimera.

The $100\%$ success rate is also maintained for the data augmentation version, indicating that while it does not, and indeed could not, further improve the success rate, it is not detrimental. This stands in contrast to what observed for Chimera where its impact was sometimes positive and other times negative. 

The results clearly show that the RL agent is very effective for this type of setting.

\paragraph{Number of Qubits}
By comparing the number of qubits required for a successful minor embedding we can see how it remains low for the base version of the agent when minor embedding graphs on small $H_{size}\leq 4$. For $H_{size}=4$ a graph $|G|=4$ requires only 7 qubits, while a graph $|G|=6$ requires 13 and finally $|G|=8$ requires 22. While the number of qubits required increases faster than the size of the graph $G$, this increase is limited. 
If we consider small $|G|\leq 4$ we can see how the number of qubits required remains limited as the size of $H$ increases, \idest for $|G|=4$ the minor embedding requires 7 qubits on $H_{size}=4$, 13 on $H_{size}=6$, and 21 on $H_{size}=8$. So again increasing the size of $H$ results in a minor embedding that is less efficient, which can be explained by the difficulty of the agent to correctly model the larger hardware graph and possibly by the detrimental effect caused by the increased action space the agent needs to explore.

By increasing both $G$ and $H$ we see that the minor embedding starts to require a very large number of qubits beyond $H_{size}=|G|=8$, with a minor of almost 750 qubits, again suggesting that the MLP-based agent struggles when the topology of the graph becomes complex and, despite being able to provide a successful minor embedding, this comes with an inefficient use of the available qubits.

By looking at the effect of the data augmentation we can see a consistent increase in the number of qubits required. This difference depends on the scenario. For small $|G|\leq 4$ the data augmentation requires a similar number of qubits until $H_{size} \geq 7$, where it starts to increase steeply. See for example $|G|=4$ where its minor embedding requires 13 qubits with the base variant and 21 with the data augmentation one when $H_{size}=6$, while it requires 15 with the base version and a much larger 96 with the data augmentation one for the larger $H_{size}=7$. The difference becomes even more marked for $H_{size}=8$. This again confirms that the architecture of the MLP-based agent struggles with modelling minor embeddings on larger graphs and the data augmentation strains the model ability further.
A similar observation is valid also by restricting to lower $H_{size} \leq 5$ and increasing the size of $G$. With $H_{size} = 5$ the minor embedding of $|G|=4$ requires 10 qubits with the base agent and 9 with the data augmentation one. This rare instance where the data augmentation requires fewer qubits can be explained with its very high variance in many experiments. If however we move to larger $G$ the discrepancy again increases in favour of the base model. For $|G|=6$ the base model requires 26 qubits while the data augmentation version requires 88 and finally for $|G|=8$ the base version requires 45 qubits while the data augmentation around 400, almost ten times as many.
A curious pattern is that as both $H$ and $G$ grow past $7$, the discrepancy reduces substantially. For the larger experiment where $H_{size}=|G|=8$ the base version requires almost 750 qubits while the data augmentation version around 1000. A possible explanation for this effect is that in this setting the base agent too needs to explore a much larger portion of the graph $H$ and so it has to learn a more difficult problem, whereas for smaller ones the base model can solve a simpler problem compared to the agent that uses data augmentation.

\begin{figure}
    \centering
    \includegraphics[width=0.5\linewidth]{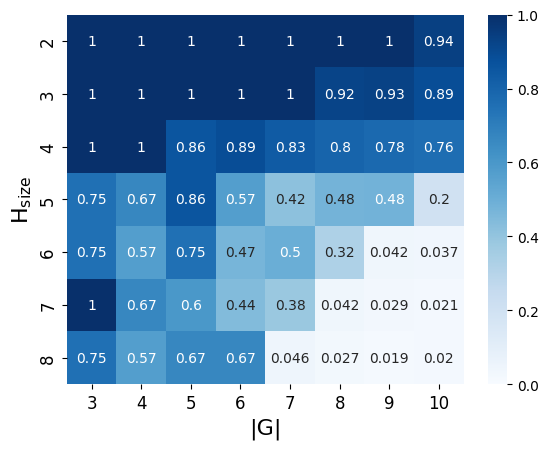}
    \caption{Qubit Efficiency Ratio for the fully connected scenario on Zephyr.}
    \label{fig:rl:mlp:zephyr_qur}
\end{figure}

\figref{fig:rl:mlp:zephyr_qur} shows a visualization of the Qubit Efficiency Ratio. 
For almost all the sizes of $G$ embedded on $H_{size} \leq 4$ (around 500 qubits) the agent is able to use a number of qubits that is either identical or rather close to what is required by \texttt{minorminer}. This is a similar finding as in Chimera and again is due to how both \texttt{minorminer} and the agent are able to find minor embeddings that are either optimal or very close to optimal.

A similar observation can be made if we compare the number of qubits for small $|G| \leq 4$ embedded on increasingly larger hardware graphs. There the minor embedding is somewhat less efficient with \texttt{minorminer} being able to use two-thirds of the qubits required by the RL agent. The smaller number of qubits required by the minor embedding can be attributed to the higher connectivity of Zephyr which results in shorter chains and therefore requires a less complex modelling of the embedded graph topology.

However, when embedding graphs with larger $|G|$ and $H_{size}$ the minor embedding becomes very inefficient. For example, for $|G|=9$ and $H_{size}=6$ the minor produced by the agent requires around 25 times as many qubits as \texttt{minorminer}. Again, the impact of this effect is relatively small as the results show that those graphs can be easily and efficiently embedded in smaller $H$ graphs and so, one can use smaller ones as target and move to larger ones only if the minor embedding fails. This mitigation strategy would alleviate this problem until graph $G$  is large enough to require a larger $H$ as well.

\begin{figure}[!h]
    \centering
    \includegraphics[width=\linewidth]{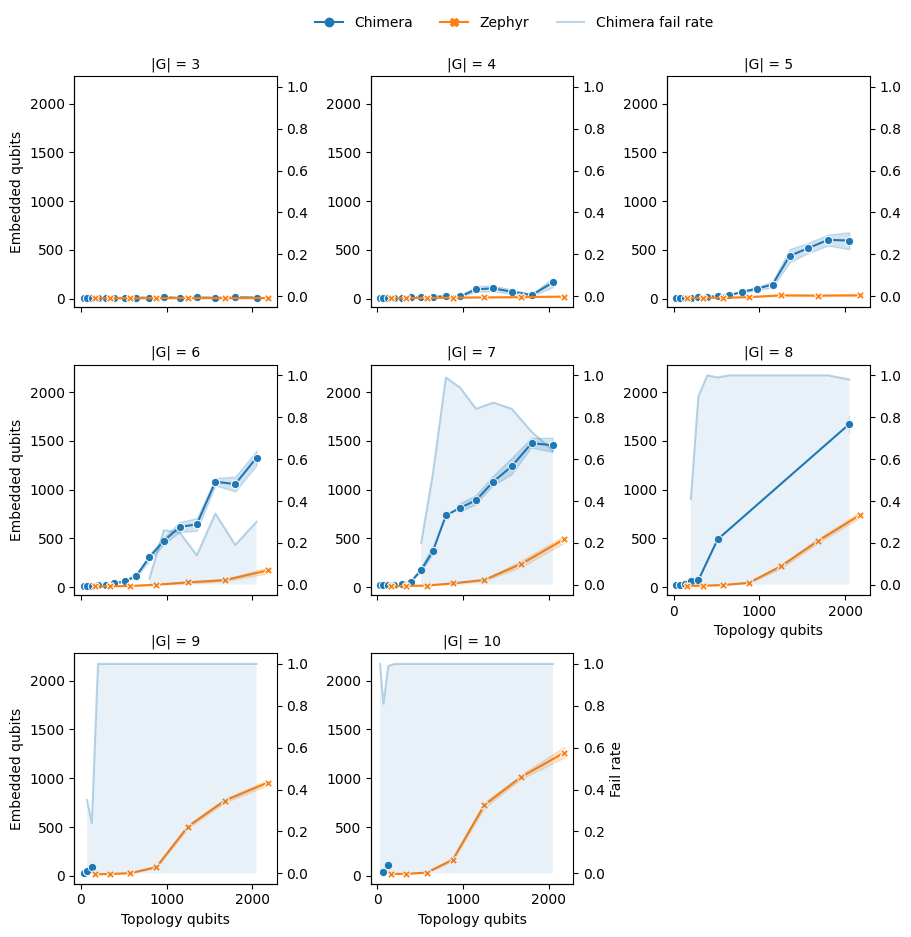}
    \caption{Number of qubits used by the RL agent to embed fully connected graphs of varying sizes $|G|$ onto Chimera and Zephyr topologies with increasing numbers of hardware qubits (to allow a comparison between the two hardware topologies).
    Each subplot corresponds to a fixed $|G|$, showing the qubit usage for both topologies, along with the minor embedding failure rate on Chimera (on the right $y$-axis, shaded area).
    While both topologies perform similarly for small $|G|$, Chimera exhibits higher qubit usage and increasing failure rates for larger graphs.
    Zephyr consistently achieves successful minor embeddings with fewer qubits and no observed failures across all tested scenarios.}
    \label{fig:rl:mlp:line}
\end{figure}

\subsubsection{Topology Impact on Minor Embedding}
\figref{fig:rl:mlp:line} shows a direct comparison of how the agent behaves when embedding graphs of increasing size onto Chimera and Zephyr topologies. The most important difference between the two topologies is the connectivity, which for Chimera is 6 while for Zephyr is of 20. As such, we expect that, given the same $G$, the minor embedding's size on Zephyr will be much smaller than on Chimera and the chains much shorter. This will result in an easier minor embedding especially as the size of $G$ increases, which is indeed what we can see.
For smaller graphs, particularly when $|G|$ is 3, 4, or 5, both Chimera and Zephyr perform similarly, with low qubit usage and no noticeable failures.  
However, as the the size of $G$ increases beyond this point, some trends start to emerge.

In the case of Chimera, the number of qubits required grows more quickly, especially from $|G| = 6$ onwards.  
Along with the rise in qubit usage, the failure rate also starts to increase and becomes prominent from $|G| = 7$ to $|G| = 10$.  
For the largest graphs, the failure rate in Chimera reaches high levels, and in many cases, minor embeddings are no longer consistently successful.  
This confirms how the minor embedding process becomes more difficult or less reliable as the complexity of the problem graph increases.

Zephyr, on the other hand, shows a different pattern.  
The qubit usage grows more slowly with graph size and remains lower than in Chimera for the same $|G|$ and number of qubits.  
More notably, the RL agent does not encounter failures when embedding into Zephyr, even as the problem size grows.  
This more stable behaviour indicates that the model can reliably find minor embeddings across all tested graph sizes when working with the Zephyr topology.  

Overall, while both topologies appear sufficient for smaller graphs, Zephyr tends to support minor embeddings more efficiently and consistently as graph complexity increases. In contrast, ME on Chimera is more challenging and the agent shows signs of strain both in terms of resource requirements and reliability when faced with larger or more connected graphs. 
This is likely due to the higher connectivity offered by the Zephyr topology \wrt to Chimera, resulting in easier ways for the agent to connect different chains in the minor. Based on this, the limitations of the agent's architecture are effectively mitigated by the improvements in hardware connectivity.

\subsection{Randomly Generated Problem Graphs}
\begin{figure}
    \centering
    \begin{subfigure}[b]{0.48\textwidth}
        \centering
        \includegraphics[width=\textwidth]{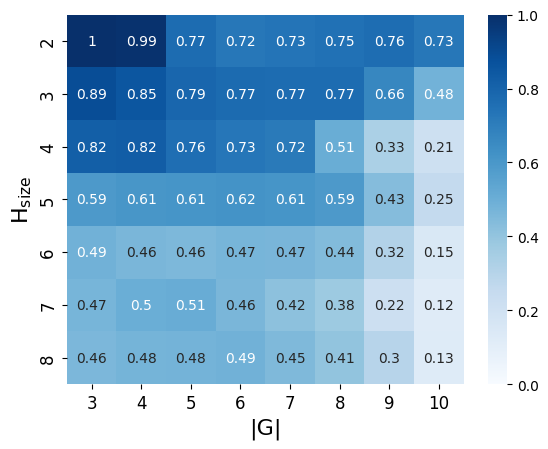}
        \caption{Average QER}
        \label{fig:rl:mlp:zephyr_dataset_qur_avg}
    \end{subfigure}
    \hfill
    \begin{subfigure}[b]{0.48\textwidth}
         \centering
         \includegraphics[width=\textwidth]{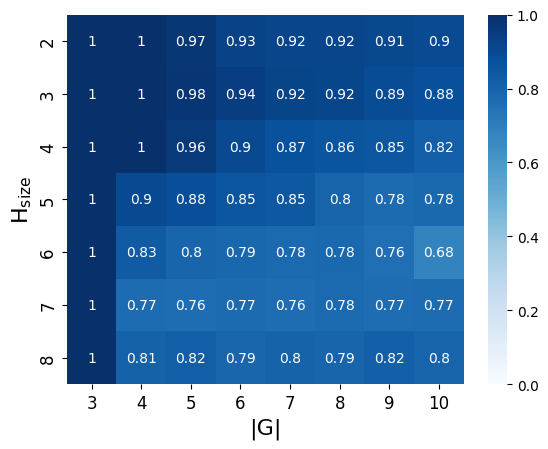}
         \caption{Best QER}
         \label{fig:rl:mlp:zephyr_dataset_qur_best}
    \end{subfigure}
    \caption{Qubit Efficiency Ratio for the random graphs scenario on Zephyr.}
    \label{fig:rl:mlp:zephyr_dataset_qur}
\end{figure}

The previous scenario focused on the minor embedding of a fully connected $G$, which constitutes a particularly challenging problem where the embedded graph will be very large and complex. This second scenario aims to test the effectiveness of the RL agent on the randomly generated problem graphs described in Section \ref{sec:rl4me:dataset}. Given that Zephyr is the most recent architecture, we only use this topology for this experiment. The results are reported in \figref{fig:rl:mlp:zephyr_dataset_qur}.

From a high level the two scenarios (fully connected and random graphs) exhibit a similar overall pattern, with an optimal or near-optimal minor embedding for small $G$ and small $H$. The differences become much more pronounced as both sizes increase.
If we look at the largest experiment, with $H_{size}=8$ and $|G|=10$ we can see that the best qubit efficiency ratio is $0.8$ indicating how the agent found a very efficient minor embedding, only slightly larger in terms of number of qubits than \texttt{minorminer}. However, if we look at the average one, we see it is $0.13$ indicating that most minor embeddings require more than ten times as many qubits as \texttt{minorminer}. This indicates that the MLP-based agent is reaching again the limits of its ability to model the structure of the graph.
When $G$ was fully connected this value was $0.02$ indicating that the agent required \emph{fifty times} as many qubits as \texttt{minorminer}, which can be expected given that the fully connected graphs have a more complex topology to embed.
These results indicate that when the agent is tasked to embed a graph that is less connected, it behaves much better.
Again, this points to the challenging task of modelling effectively the topology of both graphs as well as of the incomplete minor embedding that is being built by the RL agent, to decide which should be the most effective action.
Furthermore, the experiments on the random problem graphs do not show the sharp increase in the number of qubits that was present in fully connected graphs, where in some scenarios the number of qubits jumped by a factor of 10 by simply increasing either $|G|$ or $H_{size}$ by 1.
For example, with $H_{size}=|G|=7$, the QER for the minor embedding of the fully connected $G$ was $0.38$ while the slightly larger $|G|=8$ had $0.04$.
This points to a crucial threshold where the modelling becomes very ineffective.
This threshold does not appear to be present for the minor embedding of random problem graphs, likely due to how they are less connected and the resulting minor is simpler, pushing this crucial threshold towards larger $H$ and $G$. However, the large difference between the average and best QER indicates that different runs and graphs have a very high variance and therefore, even though the best QER is good, the training is becoming unstable in those regions.

\begin{table}
\caption{Comparison of the number of qubits required for the minor embedding of random graphs built by the RL agent, when using data augmentations during training only, and using them both during training and testing. The \textbf{DA} column indicates whether data augmentation was used only during training (Train) or during both training and testing (Test).}
\label{tab:rl:mlp:da}
\centering
\footnotesize
\input{mlp_dataset_da}
\end{table}

However, additional experiments were conducted using the data augmentation mechanism presented in \secref{sec:model:da}.
Results are presented in \tabref{tab:rl:mlp:da} and show two different sets of experiments, one is using data augmentations only during training, testing the agent on the same H graph for all action steps, while the other uses data augmentations also during the testing phase.
As we can see, while testing on a fixed H graph results in inefficient minor embedding and seems to confuse the agent, as it happened in the fully connected scenario, while testing on augmented H graphs, with a different permutation of the original graph at each action step, is very effective and helps the agent to find a substantially smaller minor embedding. For example, when $H_{size}=|G|=4$ both versions produce minor embeddings requiring 5 qubits, when $H_{size}=|G|=6$ using data augmentation during testing produces minors of 10 qubits while not using it increases the minor's size to 47, and when $H_{size}=|G|=8$ this difference becomes even more pronounced with a minor embedding requiring 18 qubits if using augmentations during testing and 317 if not.

\begin{figure}
    \centering
    \begin{subfigure}[b]{0.48\textwidth}
        \centering
        \includegraphics[width=\textwidth]{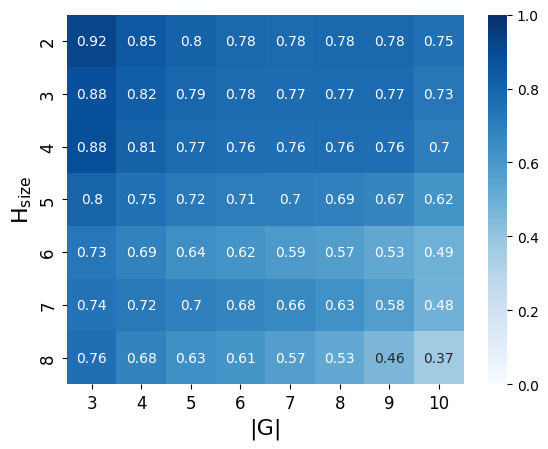}
        \caption{Average QER}
        \label{fig:rl:mlp:zephyr_dataset_da_qur_avg}
    \end{subfigure}
    \hfill
    \begin{subfigure}[b]{0.48\textwidth}
         \centering
         \includegraphics[width=\textwidth]{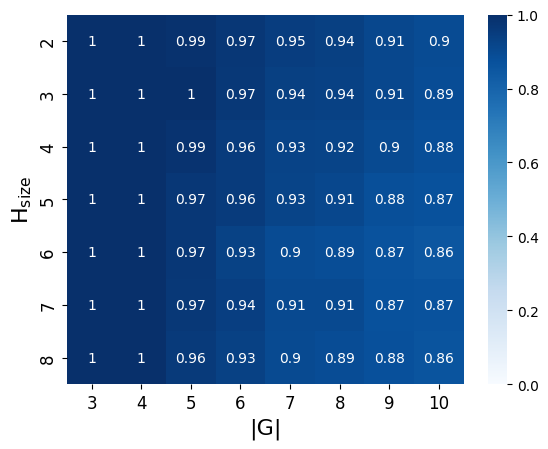}
         \caption{Best QER}
         \label{fig:rl:mlp:zephyr_dataset_da_qur_best}
    \end{subfigure}
    
    \caption{Qubit Efficiency Ratio for the random graphs scenario on Zephyr when training with data augmentation.}
    \label{fig:rl:mlp:zephyr_dataset_da_qur}
\end{figure}

Average and best QER are shown in \figref{fig:rl:mlp:zephyr_dataset_da_qur}.
While for low H and G sizes ($H_{size}=2,3$, $|G|=3,4$) the average QER is lower than the base version of the agent, the overall results show a clear improvement when using data augmentation both in training and testing.
Similar tests, using the data augmentation mechanism also in the testing phase, were conducted on the fully connected problems, but did not show any relevant difference.
This may be due to the regular structure of the problem graphs being learned.

Overall, these results indicate that while the data augmentation strategies did not appear effective on fully connected graphs, it is very useful when applied both during training and testing on randomly generated graphs.

\subsection{Training Stability and Convergence}
\label{sec:training_convergence}
Training \ac{RL} agents is a notoriously challenging task due to the inherent complexity of the underlying optimization landscape. Unlike supervised learning, where the objective function is typically convex or at least stationary, RL loss surfaces are often \emph{non-convex}, \emph{non-smooth}, and \emph{non-stationary}. The agent's actions influence its environment, which in turn modifies future observations and rewards, yielding a  moving target for the optimizer. Furthermore, in actor-critic setups such as the one employed in this work, the interaction between the policy and the value function can lead to phenomena like \emph{policy collapse}, \emph{vanishing gradients}, or \emph{catastrophic forgetting}, especially when rewards are sparse or delayed. As a consequence, convergence is not guaranteed, and instabilities are often observed in training dynamics. Monitoring learning curves over long training horizons becomes essential to assess whether the agent is progressively improving or being misled by spurious feedback.

In this work, the reward function was designed to be both dense and aligned with the minor embedding objective: the agent receives a fixed negative reward at each step. This formulation encourages the agent to minimize the number of actions taken to construct a valid minor, thereby promoting the formation of shorter chains. The use of immediate, consistent penalties avoids the difficulties associated with sparse or delayed rewards. While alternative rewards such as energy-based or sparse terminal rewards could be explored in future work, this simple step-based penalty shows to be sufficiently stable and effective.

\begin{figure}
    \centering
    \includegraphics[width=\linewidth]{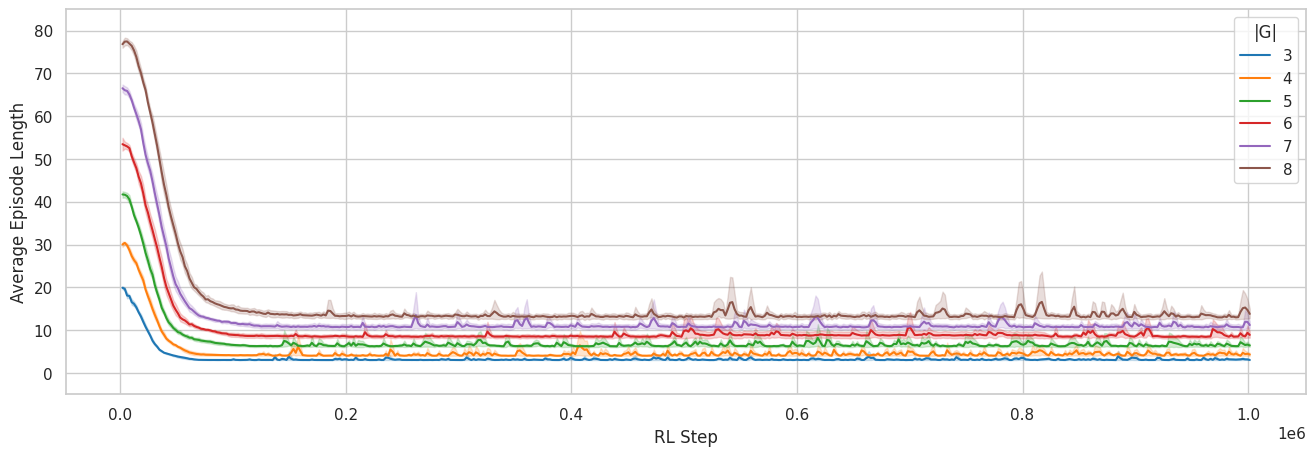}
    \caption{Convergence of the RL agent training on Zephyr with $H_{size} = 2$ in the fully connected graphs scenario.}
    \label{fig:rl:mlp:ep_len_h2}
\end{figure}

\begin{figure}
    \centering
    \includegraphics[width=\linewidth]{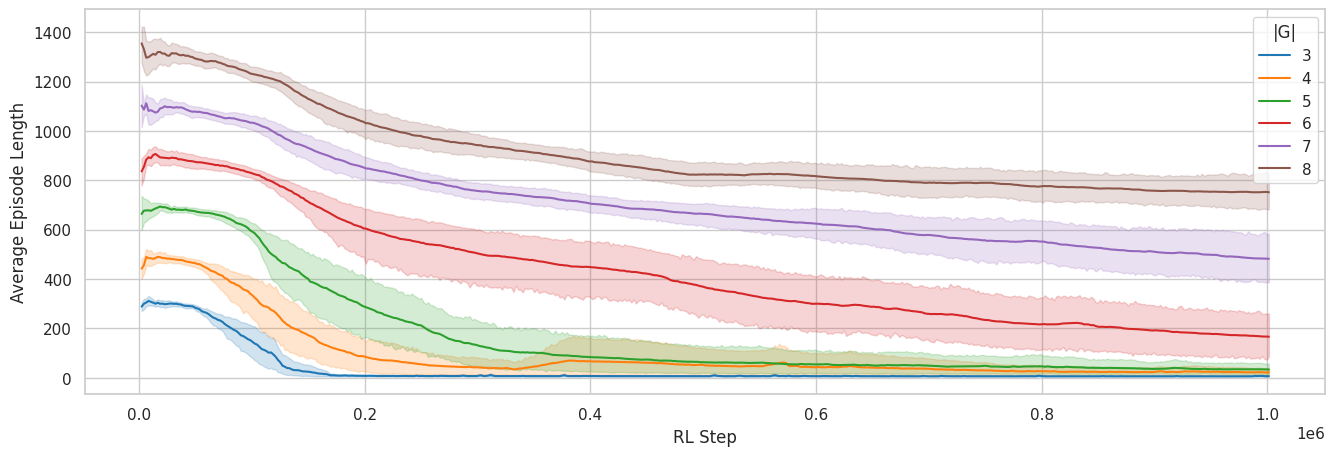}
    \caption{Convergence of the RL agent training on Zephyr with $H_{size} = 8$ in the fully connected graphs scenario.}
    \label{fig:rl:mlp:ep_len_h8}
\end{figure}

Figures~\ref{fig:rl:mlp:ep_len_h2} and~\ref{fig:rl:mlp:ep_len_h8} show the convergence behaviour of the RL agent trained on the same fully connected problem graph but targeting two different Zephyr graphs, respectively with $H_{size}=2$ and $H_{size}=8$ unit cells. Each curve represents the mean and standard deviation of 10 runs of problem graphs $G$ of different sizes, reporting the mean episode length over episode batches. Episode length corresponds to the number of actions taken by the agent before the episode terminates, either by successfully constructing a complete minor embedding or by failing to produce a valid one.

In this context, shorter episode lengths typically indicate more efficient minor embedding policies because the agent finds a solution in fewer steps. Hence, improvements in the agent can be inferred from a consistently declining episode length that stabilizes at a value that should be the optimal number of qubits required for the minor embedding.

In the $H_{size} = 2$ setting (Figure~\ref{fig:rl:mlp:ep_len_h2}), corresponding to a relatively small hardware topology (160 qubits), the agent displays robust convergence across all $G$ sizes and random initializations. Episode lengths drop rapidly during the early phase of training (first 100k steps), stabilizing to relatively low values. This suggests that the agent learns a consistent policy for constructing valid minors in a compact search space.

In contrast, the $H_{size} = 8$ setting (Figure~\ref{fig:rl:mlp:ep_len_h8}) involves a significantly larger and more complex hardware graph (2176 qubits), resulting in longer episodes and a more challenging minor embedding task. Nonetheless, the majority of $G$ show clear evidence of learning, episode lengths initially increase or plateau, possibly due to exploration noise or suboptimal early policies, before steadily decreasing over the course of training. The final episode lengths vary across runs, indicating some residual variance, but in all successful runs the trend is downward. One or two trajectories for higher $G$ sizes stagnate at higher values, reflecting  partial convergence or suboptimal local minima, again consistent with known RL difficulties. Overall, however, the agent exhibits promising training dynamics even in a high-dimensional action space, indicating that the policy architecture and training setup support meaningful learning.

Taken together, these results suggest that the agent is capable of learning effective minor embedding policies for moderately sized Zephyr topologies. While occasional seed failures or plateaus are to be expected in this setting, the general trend across both experiments demonstrates that the agent is able to cope with increasing hardware complexity without suffering from severe instability or divergence.

\section{Conclusion}

In this work we proposed a Reinforcement Learning agent trained with Proximal Policy Optimization to tackle the problem of Minor Embedding, with a goal of exploring machine learning methods that would offer a higher degree of flexibility compared to traditional heuristics.

The MLP-based reinforcement learning agent demonstrated the ability to learn meaningful minor embedding strategies but only for relatively small graphs and hardware topologies.
The success rate of the agent rapidly decreases for larger graphs on the older Chimera topology and the number of qubits required remained higher than \texttt{minorminer}, sometimes significantly. In contrast, on the newer Zephyr topology the agent was able to benefit from the higher qubit connectivity exhibiting a much higher success rate.

The use of data augmentations in the form of random graph permutations proved beneficial for randomly generated graphs when used both during the training and testing phases, resulting in a considerable reduction of the number of qubits required by the minor embedding.

Overall, while our results indicate that the use of RL agents for minor embedding is possible, they also point to limitations of the agent architecture which struggles to model the topology of the graphs, limiting its ability to fully exploit structural regularities in the problem and hardware graphs. Based on this, an important future direction is to explore the use of other architectures such as Graph Neural Networks (GNNs) which natively model the characteristics of hardware structures in a way that could make the training more efficient and the agent more robust.

\bmhead{Acknowledgements}
We acknowledge the financial support from ICSC - ``National Research Centre in High Performance Computing, Big Data and Quantum Computing'', funded by European Union – NextGenerationEU.

\clearpage
\appendix
\section{Full Results}
\label{app:results}

This section reports the full results for the minor embedding of fully connected graphs on the Chimera topology (see \tabref{table:rl:mlp:chimera:g3-6} for $|G|=3$ to $6$, and \tabref{table:rl:mlp:chimera:g7-10} for $|G|=7$ to $10$) and the Zephyr topology (see \tabref{table:rl:mlp:zephyr:g3-6} for $|G|=3$ to $6$, and \tabref{table:rl:mlp:zephyr:g7-10} for $|G|=7$ to $10$).

\begin{table}[h!]
\centering
\caption{Success rate, as well as average and standard deviation of the number of qubits required by the minor embedding built by the RL agent on fully connected problem graphs G (from $|G|=3$ to $6$) mapped to Chimera topology graphs H.}
\label{table:rl:mlp:chimera:g3-6}
\footnotesize
\input{app_mlp_chimera_1}
\end{table}

\begin{table}[h!]
\centering
\caption{Success rate, as well as average and standard deviation of the number of qubits required by the minor embedding built by the RL agent on fully connected problem graphs G (from $|G|=7$ to $10$) mapped to Chimera topology graphs H.}
\label{table:rl:mlp:chimera:g7-10}
\footnotesize
\input{app_mlp_chimera_2}
\end{table}

\begin{table}[h!]
\centering
\caption{Success rate, as well as average and standard deviation of the number of qubits required by the minor embedding built by the RL agent on fully connected problem graphs G (from $|G|=3$ to $6$) mapped to Zephyr topology graphs H.}
\label{table:rl:mlp:zephyr:g3-6}
\footnotesize
\input{app_mlp_zephyr_1}
\end{table}

\begin{table}[h!]
\centering
\caption{Success rate, as well as average and standard deviation of the number of qubits required by the minor embedding built by the RL agent on fully connected problem graphs G (from $|G|=7$ to $10$) mapped to Zephyr topology graphs H.}
\label{table:rl:mlp:zephyr:g7-10}
\footnotesize
\input{app_mlp_zephyr_2}
\end{table}


\clearpage
\bibliography{bibliography}

\end{document}

%% file: acronyms.tex
\acrodef{CL}[CL]{Curriculum Learning}
\acrodef{IAM}[IAM]{Invalid Action Masking}
\acrodef{ME}[ME]{Minor Embedding}
\acrodef{ML}[ML]{Machine Learning}
\acrodef{PPO}[PPO]{Proximal Policy Optimization}
\acrodef{RL}[RL]{Reinforcement Learning}
\acrodef{QA}[QA]{Quantum Annealing}
\acrodef{QPU}[QPU]{Quantum Processing Unit}
\acrodef{QUBO}[QUBO]{Quadratic Unconstrained Binary Optimization}

%% file: mlp_chimera.tex
\begin{tabular}{c|c|cc|cc|cc|cc}
\toprule
\multirow{2}{*}{$\mathbf{H_{size}}$} & \multirow{2}{*}{\textbf{DA}} & \multicolumn{2}{c}{$\mathbf{|G| = 4}$} & \multicolumn{2}{c}{$\mathbf{|G| = 6}$} & \multicolumn{2}{c}{$\mathbf{|G| = 8}$} & \multicolumn{2}{c}{$\mathbf{|G| = 10}$} \\
 &  & SR\% & \#Q & SR\% & \#Q & SR\% & \#Q & SR\% & \#Q \\
\midrule
\multirow{2}{*}{2}  &        & 100 & $7\pm3$   & 100 & $14\pm0$   & 100 & $23\pm2$   & 0   & --- \\
 & \checkmark & 100 & $6\pm1$   & 100 & $14\pm0$   & 98  & $23\pm2$   & 0   & --- \\
\midrule
\multirow{2}{*}{4}  &      & 100 & $6\pm1$   & 100 & $14\pm1$   & 100 & $29\pm6$   & 1   & $115\pm0$ \\
 & \checkmark & 100 & $7\pm2$   & 100 & $17\pm9$   & 100 & $42\pm20$  & 0   & --- \\
\midrule
\multirow{2}{*}{6}  &      & 100 & $9\pm2$   & 100 & $22\pm5$   & 10  & $73\pm2$   & 0   & --- \\
 & \checkmark & 100 & $10\pm11$ & 100 & $64\pm62$  & 24  & $257\pm18$ & 0   & --- \\
\midrule
\multirow{2}{*}{8}  &      & 100 & $15\pm5$  & 100 & $61\pm25$  & 1   & $490\pm0$  & 0   & --- \\
 & \checkmark & 100 & $32\pm37$ & 97  & $206\pm149$ & 5   & $439\pm51$ & 0   & --- \\
\midrule
\multirow{2}{*}{10} &       & 100 & $23\pm8$  & 97  & $308\pm176$ & 0   & ---        & 0   & --- \\
 & \checkmark & 100 & $119\pm126$ & 77  & $593\pm148$ & 0   & ---        & 0   & --- \\
\midrule
\multirow{2}{*}{12} &       & 100 & $96\pm148$ & 75  & $617\pm231$ & 0   & ---        & --- & --- \\
 & \checkmark & 100 & $294\pm161$ & 69  & $961\pm138$ & 0   & ---        & --- & --- \\
\midrule
\multirow{2}{*}{14} &       & 100 & $70\pm116$ & 66  & $1081\pm158$ & 0   & ---        & --- & --- \\
 & \checkmark & 100 & $412\pm212$ & 69  & $1323\pm166$ & 1   & $1182\pm0$ & --- & --- \\
\midrule
\multirow{2}{*}{16} &       & 100 & $167\pm251$ & 70  & $1328\pm314$ & 2   & $1674\pm129$ & --- & --- \\
 & \checkmark & 100 & $642\pm292$ & 50  & $1451\pm292$ & 0   & ---        & --- & --- \\
\bottomrule
\end{tabular}

%% file: mlp_zephyr.tex
\begin{tabular}
{c|c|cc|cc|cc|cc}
\toprule
\multirow{2}{*}{$\mathbf{H_{size}}$} & \multirow{2}{*}{\textbf{DA}} & \multicolumn{2}{c}{$\mathbf{|G| = 4}$} & \multicolumn{2}{c}{$\mathbf{|G| = 6}$} & \multicolumn{2}{c}{$\mathbf{|G| = 8}$} & \multicolumn{2}{c}{$\mathbf{|G| = 10}$} \\
& & SR\%  & \#Q & SR\%  & \#Q & SR\%  & \#Q & SR\%  & \#Q \\
\midrule
\multirow{2}{*}{2}
 & & 100 & $4\pm0$ & 100 & $9\pm1$ & 100 & $14\pm6$ & 100 & $18\pm1$ \\
 & \checkmark & 100 & $4\pm0$ & 100 & $9\pm1$ & 100 & $18\pm11$ & 100 & $23\pm11$ \\
\midrule
\multirow{2}{*}{3}
 & & 100 & $5\pm3$ & 100 & $9\pm2$ & 100 & $15\pm3$ & 100 & $20\pm3$ \\
 & \checkmark & 100 & $5\pm5$ & 100 & $12\pm9$ & 100 & $40\pm28$ & 100 & $105\pm56$ \\
\midrule
\multirow{2}{*}{4}
 & & 100 & $7\pm4$ & 100 & $13\pm2$ & 100 & $22\pm7$ & 100 & $32\pm20$ \\
 & \checkmark & 100 & $6\pm2$ & 100 & $29\pm27$ & 100 & $158\pm98$ & 100 & $319\pm71$ \\
\midrule
\multirow{2}{*}{5}
 & & 100 & $10\pm5$ & 100 & $26\pm8$ & 100 & $45\pm17$ & 100 & $165\pm72$ \\
 & \checkmark & 100 & $9\pm19$ & 100 & $88\pm72$ & 100 & $394\pm111$ & 100 & $532\pm141$ \\
\midrule
\multirow{2}{*}{6}
 & & 100 & $13\pm7$ & 100 & $49\pm49$ & 100 & $213\pm102$ & 100 & $724\pm135$ \\
 & \checkmark & 100 & $21\pm32$ & 100 & $307\pm189$ & 100 & $540\pm181$ & 100 & $861\pm148$ \\
\midrule
\multirow{2}{*}{7}
 & & 100 & $15\pm11$ & 100 & $72\pm63$ & 100 & $473\pm117$ & 100 & $1014\pm111$ \\
 & \checkmark & 100 & $96\pm125$ & 100 & $434\pm167$ & 100 & $921\pm190$ & 100 & $1093\pm195$ \\
\midrule
\multirow{2}{*}{8}
 & & 100 & $21\pm24$ & 100 & $172\pm159$ & 100 & $742\pm144$ & 100 & $1260\pm285$ \\
 & \checkmark & 100 & $169\pm163$ & 100 & $594\pm195$ & 100 & $1040\pm244$ & 99 & $1459\pm316$ \\
\bottomrule
\end{tabular}

%% file: mlp_dataset_da.tex
\begin{tabular}{c|c|cccccccc}
    \toprule
    $\mathbf{H_{size}}$ & \textbf{DA} & $\mathbf{|G| = 3}$ & $\mathbf{|G| = 4}$ & $\mathbf{|G| = 5}$ & $\mathbf{|G| = 6}$ & $\mathbf{|G| = 7}$ & $\mathbf{|G| = 8}$ & $\mathbf{|G| = 9}$ & $\mathbf{|G| = 10}$ \\
    \midrule
    \multirow{2}{*}{2} & Train & $3\pm0$ & $5\pm1$ & $6\pm1$ & $8\pm1$ & $10\pm1$ & $12\pm2$ & $14\pm4$ & $19\pm9$ \\
                       & Test & $3\pm0$ & $5\pm1$ & $6\pm1$ & $8\pm1$ & $10\pm1$ & $12\pm1$ & $14\pm2$ & $16\pm3$ \\
    \midrule
    \multirow{2}{*}{3} & Train & $3\pm0$ & $5\pm1$ & $6\pm1$ & $8\pm2$ & $11\pm3$ & $19\pm13$ & $37\pm25$ & $64\pm36$ \\
                       & Test & $3\pm1$ & $5\pm1$ & $6\pm1$ & $8\pm1$ & $10\pm1$ & $12\pm2$ & $14\pm2$ & $17\pm5$ \\
    \midrule
    \multirow{2}{*}{4} & Train & $3\pm1$ & $5\pm1$ & $7\pm1$ & $11\pm6$ & $29\pm19$ & $54\pm31$ & $89\pm44$ & $129\pm59$ \\
                       & Test & $3\pm1$ & $5\pm1$ & $7\pm1$ & $8\pm1$ & $10\pm1$ & $12\pm2$ & $14\pm3$ & $18\pm6$ \\
    \midrule
    \multirow{2}{*}{5} & Train & $4\pm1$ & $7\pm7$ & $11\pm13$ & $27\pm29$ & $54\pm45$ & $95\pm63$ & $150\pm87$ & $208\pm115$ \\
                       & Test & $4\pm1$ & $5\pm1$ & $7\pm1$ & $9\pm2$ & $11\pm3$ & $13\pm5$ & $16\pm7$ & $20\pm11$ \\
    \midrule
    \multirow{2}{*}{6} & Train & $5\pm3$ & $9\pm10$ & $16\pm20$ & $47\pm44$ & $80\pm69$ & $114\pm92$ & $156\pm119$ & $202\pm146$ \\
                       & Test & $4\pm1$ & $6\pm1$ & $8\pm4$ & $10\pm6$ & $13\pm8$ & $16\pm11$ & $20\pm17$ & $25\pm22$ \\
    \midrule
    \multirow{2}{*}{7} & Train & $4\pm1$ & $8\pm8$ & $32\pm39$ & $85\pm68$ & $149\pm94$ & $216\pm131$ & $304\pm173$ & $400\pm226$ \\
                       & Test & $4\pm1$ & $6\pm1$ & $7\pm2$ & $9\pm5$ & $12\pm7$ & $15\pm12$ & $19\pm19$ & $26\pm32$ \\
    \midrule
    \multirow{2}{*}{8} & Train & $4\pm1$ & $20\pm39$ & $53\pm85$ & $131\pm132$ & $224\pm179$ & $317\pm220$ & $437\pm266$ & $556\pm316$ \\
                       & Test & $4\pm1$ & $6\pm2$ & $8\pm5$ & $10\pm9$ & $14\pm16$ & $18\pm24$ & $24\pm42$ & $36\pm68$ \\
    \bottomrule
\end{tabular}

%% file: app_mlp_chimera_1.tex
\begin{tabular}{c|c|cc|cc|cc|cc}
\toprule
\multirow{2}{*}{$\mathbf{H_{size}}$} & \multirow{2}{*}{\textbf{DA}} & \multicolumn{2}{c}{$\mathbf{|G| = 3}$} & \multicolumn{2}{c}{$\mathbf{|G| = 4}$} & \multicolumn{2}{c}{$\mathbf{|G| = 5}$} & \multicolumn{2}{c}{$\mathbf{|G| = 6}$} \\
& & SR\% & \#Q & SR\% & \#Q & SR\% & \#Q & SR\% & \#Q \\
\midrule
\multirow{2}{*}{2} & & 100 & $4\pm0$ & 100 & $7\pm3$ & 100 & $8\pm0$ & 100 & $14\pm0$ \\
& \checkmark & 100 & $4\pm0$ & 100 & $6\pm1$ & 100 & $8\pm0$ & 100 & $14\pm0$ \\
\midrule
\multirow{2}{*}{3} & & 100 & $4\pm1$ & 100 & $6\pm0$ & 100 & $8\pm0$ & 100 & $14\pm2$ \\
& \checkmark & 100 & $4\pm0$ & 100 & $6\pm2$ & 100 & $8\pm0$ & 100 & $15\pm1$ \\
\midrule
\multirow{2}{*}{4} & & 100 & $5\pm2$ & 100 & $6\pm1$ & 100 & $10\pm5$ & 100 & $14\pm1$ \\
& \checkmark & 100 & $4\pm1$ & 100 & $7\pm2$ & 100 & $9\pm4$ & 100 & $17\pm9$ \\
\midrule
\multirow{2}{*}{5} & & 100 & $6\pm4$ & 100 & $7\pm1$ & 100 & $11\pm3$ & 100 & $18\pm3$ \\
& \checkmark & 100 & $5\pm5$ & 100 & $6\pm0$ & 100 & $9\pm8$ & 100 & $20\pm18$ \\
\midrule
\multirow{2}{*}{6} & & 100 & $5\pm1$ & 100 & $9\pm2$ & 100 & $19\pm11$ & 100 & $22\pm5$ \\
& \checkmark & 100 & $5\pm1$ & 100 & $10\pm11$ & 100 & $28\pm33$ & 100 & $64\pm62$ \\
\midrule
\multirow{2}{*}{7} & & 100 & $5\pm2$ & 100 & $12\pm3$ & 100 & $21\pm6$ & 100 & $38\pm21$ \\
& \checkmark & 100 & $5\pm2$ & 100 & $10\pm12$ & 100 & $51\pm71$ & 100 & $77\pm88$ \\
\midrule
\multirow{2}{*}{8} & & 100 & $8\pm3$ & 100 & $15\pm5$ & 100 & $24\pm8$ & 100 & $61\pm25$ \\
& \checkmark & 100 & $5\pm8$ & 100 & $32\pm37$ & 100 & $151\pm105$ & 97 & $206\pm149$ \\
\midrule
\multirow{2}{*}{9} & & 100 & $9\pm5$ & 100 & $19\pm8$ & 100 & $36\pm23$ & 100 & $118\pm68$ \\
& \checkmark & 100 & $6\pm10$ & 100 & $15\pm35$ & 100 & $130\pm138$ & 75 & $464\pm122$ \\
\midrule
\multirow{2}{*}{10} & & 100 & $9\pm4$ & 100 & $23\pm8$ & 100 & $70\pm39$ & 97 & $308\pm176$ \\
& \checkmark & 100 & $20\pm36$ & 100 & $119\pm126$ & 100 & $241\pm191$ & 77 & $593\pm148$ \\
\midrule
\multirow{2}{*}{11} & & 100 & $13\pm7$ & 100 & $24\pm12$ & 100 & $101\pm88$ & 74 & $476\pm149$ \\
& \checkmark & 100 & $64\pm83$ & 100 & $263\pm130$ & 97 & $502\pm180$ & 61 & $778\pm93$ \\
\midrule
\multirow{2}{*}{12} & & 100 & $7\pm3$ & 100 & $96\pm148$ & 100 & $144\pm160$ & 75 & $617\pm231$ \\
& \checkmark & 100 & $99\pm98$ & 100 & $294\pm161$ & 99 & $622\pm240$ & 69 & $961\pm138$ \\
\midrule
\multirow{2}{*}{13} & & 100 & $12\pm5$ & 100 & $106\pm152$ & 99 & $438\pm335$ & 86 & $645\pm317$ \\
& \checkmark & 100 & $220\pm113$ & 100 & $369\pm234$ & 97 & $853\pm203$ & 67 & $1048\pm191$ \\
\midrule
\multirow{2}{*}{14} & & 100 & $8\pm3$ & 100 & $70\pm116$ & 100 & $518\pm277$ & 66 & $1081\pm158$ \\
& \checkmark & 100 & $238\pm180$ & 100 & $412\pm212$ & 100 & $977\pm232$ & 69 & $1323\pm166$ \\
\midrule
\multirow{2}{*}{15} & & 100 & $13\pm10$ & 100 & $36\pm57$ & 100 & $603\pm290$ & 81 & $1057\pm362$ \\
& \checkmark & 100 & $238\pm191$ & 100 & $459\pm248$ & 100 & $1046\pm253$ & 58 & $1450\pm229$ \\
\midrule
\multirow{2}{*}{16} & & 100 & $10\pm4$ & 100 & $167\pm251$ & 100 & $596\pm417$ & 70 & $1328\pm314$ \\
& \checkmark & 100 & $319\pm207$ & 100 & $642\pm292$ & 96 & $1000\pm281$ & 50 & $1451\pm292$ \\
\bottomrule
\end{tabular}

%% file: app_mlp_chimera_2.tex
\begin{tabular}{c|c|cc|cc|cc|cc}
\toprule
\multirow{2}{*}{$\mathbf{H_{size}}$} & \multirow{2}{*}{\textbf{DA}} & \multicolumn{2}{c}{$\mathbf{|G| = 7}$} & \multicolumn{2}{c}{$\mathbf{|G| = 8}$} & \multicolumn{2}{c}{$\mathbf{|G| = 9}$} & \multicolumn{2}{c}{$\mathbf{|G| = 10}$} \\
& & SR\% & \#Q & SR\% & \#Q & SR\% & \#Q & SR\% & \#Q \\
\midrule
\multirow{2}{*}{2} & & 100 & $18\pm1$ & 100 & $23\pm2$ & 100 & $29\pm0$ & 0 & --- \\
& \checkmark & 98 & $18\pm1$ & 98 & $23\pm2$ & 76 & $29\pm1$ & 0 & --- \\
\midrule
\multirow{2}{*}{3} & & 100 & $20\pm6$ & 100 & $26\pm8$ & 65 & $49\pm11$ & 19 & $43\pm4$ \\
& \checkmark & 100 & $20\pm2$ & 100 & $25\pm3$ & 76 & $53\pm9$ & 2 & $69\pm3$ \\
\midrule
\multirow{2}{*}{4} & & 100 & $20\pm2$ & 100 & $29\pm6$ & 76 & $88\pm19$ & 1 & $115\pm0$ \\
& \checkmark & 100 & $26\pm17$ & 100 & $42\pm20$ & 64 & $88\pm19$ & 0 & --- \\
\midrule
\multirow{2}{*}{5} & & 100 & $27\pm12$ & 59 & $63\pm33$ & 0 & --- & 0 & --- \\
& \checkmark & 100 & $38\pm33$ & 78 & $104\pm51$ & 5 & $167\pm22$ & 0 & --- \\
\midrule
\multirow{2}{*}{6} & & 100 & $32\pm21$ & 10 & $73\pm2$ & 0 & --- & 0 & --- \\
& \checkmark & 99 & $93\pm67$ & 24 & $257\pm18$ & 0 & --- & 0 & --- \\
\midrule
\multirow{2}{*}{7} & & 100 & $54\pm18$ & 0 & --- & 0 & --- & 0 & --- \\
& \checkmark & 83 & $176\pm108$ & 14 & $347\pm19$ & 0 & --- & 0 & --- \\
\midrule
\multirow{2}{*}{8} & & 80 & $176\pm92$ & 1 & $490\pm0$ & 0 & --- & 0 & --- \\
& \checkmark & 56 & $333\pm165$ & 5 & $439\pm51$ & 0 & --- & 0 & --- \\
\midrule
\multirow{2}{*}{9} & & 47 & $370\pm141$ & 0 & --- & 0 & --- & 0 & --- \\
& \checkmark & 48 & $558\pm50$ & 11 & $559\pm62$ & 1 & $599\pm0$ & 0 & --- \\
\midrule
\multirow{2}{*}{10} & & 1 & $736\pm0$ & 0 & --- & 0 & --- & 0 & --- \\
& \checkmark & 32 & $691\pm54$ & 0 & --- & 0 & --- & 0 & --- \\
\midrule
\multirow{2}{*}{11} & & 6 & $815\pm59$ & 0 & --- & --- & --- & --- & --- \\
& \checkmark & 28 & $830\pm59$ & 6 & $834\pm63$ & --- & --- & --- & --- \\
\midrule
\multirow{2}{*}{12} & & 16 & $892\pm100$ & 0 & --- & --- & --- & --- & --- \\
& \checkmark & 30 & $928\pm116$ & 0 & --- & --- & --- & --- & --- \\
\midrule
\multirow{2}{*}{13} & & 13 & $1082\pm86$ & 0 & --- & --- & --- & --- & --- \\
& \checkmark & 25 & $1087\pm100$ & 1 & $1216\pm0$ & --- & --- & --- & --- \\
\midrule
\multirow{2}{*}{14} & & 16 & $1240\pm180$ & 0 & --- & --- & --- & --- & --- \\
& \checkmark & 9 & $1303\pm130$ & 1 & $1182\pm0$ & --- & --- & --- & --- \\
\midrule
\multirow{2}{*}{15} & & 27 & $1477\pm134$ & 0 & --- & --- & --- & --- & --- \\
& \checkmark & 12 & $1466\pm124$ & 0 & --- & --- & --- & --- & --- \\
\midrule
\multirow{2}{*}{16} & & 36 & $1455\pm234$ & 2 & $1674\pm129$ & --- & --- & --- & --- \\
& \checkmark & 10 & $1673\pm190$ & 0 & --- & --- & --- & --- & --- \\
\bottomrule
\end{tabular}

%% file: app_mlp_zephyr_1.tex
\begin{tabular}{c|c|cc|cc|cc|cc}
\toprule
\multirow{2}{*}{$\mathbf{H_{size}}$} & \multirow{2}{*}{\textbf{DA}} & \multicolumn{2}{c}{$\mathbf{|G| = 3}$}& \multicolumn{2}{c}{$\mathbf{|G| = 4}$}& \multicolumn{2}{c}{$\mathbf{|G| = 5}$}& \multicolumn{2}{c}{$\mathbf{|G| = 6}$} \\
& & SR\%  & \#Q & SR\%  & \#Q & SR\%  & \#Q & SR\%  & \#Q \\
\midrule
\multirow{2}{*}{2}
 & & 100 & $3\pm0$ & 100 & $4\pm0$ & 100 & $6\pm1$ & 100 & $9\pm1$ \\
 & \checkmark & 100 & $3\pm0$ & 100 & $4\pm0$ & 100 & $7\pm1$ & 100 & $9\pm1$ \\
\midrule
\multirow{2}{*}{3}
 & & 100 & $3\pm0$ & 100 & $5\pm3$ & 100 & $8\pm2$ & 100 & $9\pm2$ \\
 & \checkmark & 100 & $3\pm0$ & 100 & $5\pm5$ & 100 & $7\pm0$ & 100 & $12\pm9$ \\
\midrule
\multirow{2}{*}{4}
 & & 100 & $4\pm2$ & 100 & $7\pm4$ & 100 & $9\pm2$ & 100 & $13\pm2$ \\
 & \checkmark & 100 & $3\pm0$ & 100 & $6\pm2$ & 100 & $10\pm6$ & 100 & $29\pm27$ \\
\midrule
\multirow{2}{*}{5}
 & & 100 & $5\pm2$ & 100 & $10\pm5$ & 100 & $16\pm5$ & 100 & $26\pm8$ \\
 & \checkmark & 100 & $4\pm2$ & 100 & $9\pm19$ & 100 & $52\pm70$ & 100 & $88\pm72$ \\
\midrule
\multirow{2}{*}{6}
 & & 100 & $6\pm1$ & 100 & $13\pm7$ & 100 & $33\pm50$ & 100 & $49\pm49$ \\
 & \checkmark & 100 & $7\pm13$ & 100 & $21\pm32$ & 100 & $117\pm76$ & 100 & $307\pm189$ \\
\midrule
\multirow{2}{*}{7}
 & & 100 & $5\pm2$ & 100 & $15\pm11$ & 100 & $31\pm24$ & 100 & $72\pm63$ \\
 & \checkmark & 100 & $16\pm36$ & 100 & $96\pm125$ & 100 & $208\pm176$ & 100 & $434\pm167$ \\
\midrule
\multirow{2}{*}{8}
 & & 100 & $6\pm3$ & 100 & $21\pm24$ & 100 & $34\pm39$ & 100 & $172\pm159$ \\
 & \checkmark & 100 & $61\pm100$ & 100 & $169\pm163$ & 100 & $406\pm237$ & 100 & $594\pm195$ \\
\bottomrule
\end{tabular}

%% file: app_mlp_zephyr_2.tex
\begin{tabular}{c|c|cc|cc|cc|cc}
\toprule
\multirow{2}{*}{$\mathbf{H_{size}}$} & \multirow{2}{*}{\textbf{DA}} & \multicolumn{2}{c}{$\mathbf{|G| = 7}$}& \multicolumn{2}{c}{$\mathbf{|G| = 8}$}& \multicolumn{2}{c}{$\mathbf{|G| = 9}$}& \multicolumn{2}{c}{$\mathbf{|G| = 10}$} \\
& & SR\%  & \#Q & SR\%  & \#Q & SR\%  & \#Q & SR\%  & \#Q \\
\midrule
\multirow{2}{*}{2}
 & & 100 & $11\pm3$ & 100 & $14\pm6$ & 100 & $15\pm2$ & 100 & $18\pm1$ \\
 & \checkmark & 100 & $12\pm4$ & 100 & $18\pm11$ & 100 & $20\pm11$ & 100 & $23\pm11$ \\
\midrule
\multirow{2}{*}{3}
 & & 100 & $13\pm2$ & 100 & $15\pm3$ & 100 & $18\pm5$ & 100 & $20\pm3$ \\
 & \checkmark & 100 & $23\pm18$ & 100 & $40\pm28$ & 100 & $91\pm53$ & 100 & $105\pm56$ \\
\midrule
\multirow{2}{*}{4}
 & & 100 & $16\pm4$ & 100 & $22\pm7$ & 100 & $26\pm22$ & 100 & $32\pm20$ \\
 & \checkmark & 100 & $85\pm83$ & 100 & $158\pm98$ & 100 & $274\pm75$ & 100 & $319\pm71$ \\
\midrule
\multirow{2}{*}{5}
 & & 100 & $39\pm18$ & 100 & $45\pm17$ & 100 & $88\pm41$ & 100 & $165\pm72$ \\
 & \checkmark & 100 & $225\pm141$ & 100 & $394\pm111$ & 100 & $487\pm104$ & 100 & $532\pm141$ \\
\midrule
\multirow{2}{*}{6}
 & & 100 & $74\pm44$ & 100 & $213\pm102$ & 100 & $504\pm92$ & 100 & $724\pm135$ \\
 & \checkmark & 100 & $387\pm201$ & 100 & $540\pm181$ & 100 & $645\pm178$ & 100 & $861\pm148$ \\
\midrule
\multirow{2}{*}{7}
 & & 100 & $239\pm171$ & 100 & $473\pm117$ & 100 & $771\pm105$ & 100 & $1014\pm111$ \\
 & \checkmark & 100 & $655\pm222$ & 100 & $921\pm190$ & 100 & $1109\pm203$ & 100 & $1093\pm195$ \\
\midrule
\multirow{2}{*}{8}
 & & 100 & $490\pm188$ & 100 & $742\pm144$ & 100 & $954\pm140$ & 100 & $1260\pm285$ \\
 & \checkmark & 100 & $788\pm190$ & 100 & $1040\pm244$ & 100 & $1287\pm239$ & 99 & $1459\pm316$ \\
\bottomrule
\end{tabular}